\documentclass[onecolumn,showpacs,preprint,preprintnumbers,amsmath,amssymb]{revtex4}
\usepackage{amssymb}
\usepackage[dvips]{graphicx}
\begin{document}

\title{ Bose-Einstein condensation of strongly correlated electrons and phonons in cuprate superconductors}
\author{A. S. Alexandrov}

\affiliation{Department of Physics, Loughborough University,
Loughborough LE11 3TU, United Kingdom\\}

\begin{abstract} The long-range  Fr\"ohlich electron-phonon  interaction
has been identified as the most essential for pairing in
high-temperature superconductors owing to  poor screening, as is now
confirmed by optical, isotope substitution, recent  photoemission
and some other measurements. I argue that low energy physics in
cuprate superconductors is that of superlight small bipolarons,
which are real-space hole pairs dressed by phonons in doped
charge-transfer Mott insulators. They are itinerant quasiparticles
existing in the Bloch states at low temperatures as also confirmed
by continuous-time quantum Monte-Carlo algorithm (CTQMC) fully
taking into account realistic Coulomb and long-range Fr\"ohlich
interactions. Here I suggest  that a parameter-free evaluation of
$T_c$, unusual  upper critical fields, the normal state Nernst
effect, diamagnetism,
 the Hall-Lorenz numbers and giant proximity effects
 strongly support the three-dimensional (3D) Bose-Einstein condensation of mobile small bipolarons  with zero off-diagonal order parameter above the
resistive critical temperature $T_c$ at variance with phase
fluctuation scenarios of cuprates.
\end{abstract}


\maketitle

\section{Introduction: Essential pairing interaction in cuprates}

 Although high-temperature superconductivity (HTS) has not yet been
targeted as `{\it the shame and despair of theoretical physics}', -
a label attributed to low-temperature superconductivity during the
first half-century after its discovery -  controversy of current
theoretical constructions has led many researchers to say that there
is no theory of HTS and no progress in understanding the phenomenon.
A significant fraction of theoretical research in the field   has
suggested that the interaction in novel superconductors is
essentially repulsive and unretarted, and it could provide high
$T_{c}$ without  phonons. Indeed strong  onsite repulsive
correlations (Hubbard $U$) are essential in shaping the insulating
state of  undoped (parent)  compounds.  Different from conventional
band-structure insulators with completely filled and empty Bloch
bands, the Mott insulator arises from a potentially metallic
half-filled band as a result of  the Coulomb blockade of electron
tunnelling to neighboring sites \cite{mott}. However,  the Hubbard
$U$ model shares an inherent difficulty in determining the order
when the Mott-Hubbard insulator is doped. While some groups have
claimed that it describes high-$T_{c}$ superconductivity at finite
doping, other authors could not find any superconducting
instability. Therefore it has been concluded that models of this
kind are highly conflicting and confuse the issue by exaggerating
the magnetism rather than clarifying it \cite{lau}.

 Here I discuss a multi-polaron approach to the problem based on the
 bipolaronic
extension of the BCS theory to the strong-coupling regime
\cite{alebook1}. Attractive electron correlations,  prerequisite to
any superconductivity, are caused by an almost unretarted
electron-phonon (e-ph) interaction sufficient to overcome the direct
Coulomb repulsion in this regime. Low energy physics is that of
small polarons and bipolarons, which are real-space electron (hole)
pairs dressed by phonons. They are itinerant quasiparticles existing
in the Bloch states at temperatures below the characteristic phonon
frequency. Since there is almost no retardation (i.e. no
Tolmachev-Morel-Anderson logarithm) reducing the Coulomb repulsion,
e-ph interactions should be relatively strong to overcome the direct
Coulomb repulsion, so carriers \emph{must} be polaronic to form
pairs in novel superconductors.

In our approach to cuprate superconductors  we take the view that
cuprates and related transition metal oxides are charge-transfer
Mott-Hubbard insulators at $any$ relevant level of doping
\cite{alebook1}. The one-particle density-of-states (DOS) of
cuprates is schematically represented by Fig.1, as it has been
established in a number of site-selective experiments \cite{sel} and
in the first-principle numerical ("LDA+U") \cite{first} and
semi-analytical cluster \cite{ovc} band structure calculations
properly taking into account the strong on-site repulsion.

\begin{figure}[tbp]
\begin{center}
\includegraphics[angle=-90,width=0.80\textwidth]{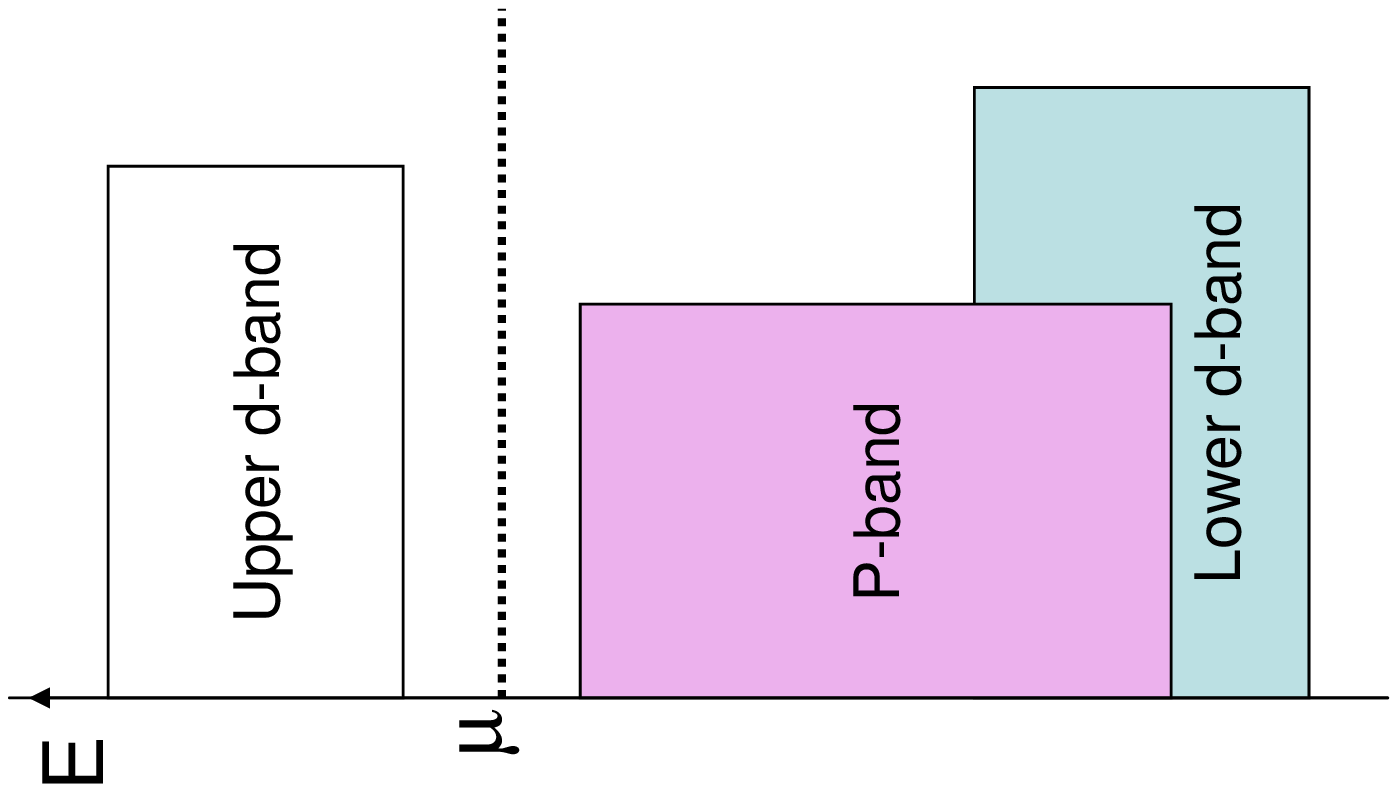}
\end{center}
\caption{ DOS in cuprates. The chemical potential $\mu$ is inside
the charge transfer gap as observed in the tunnelling experiments
\cite{boz0} because of  bipolaron formation \cite{alebook1} . It
could enter the oxygen band in overdoped cuprates, if bipolarons
 coexist with unpaired  degenerate polarons \cite{narlikar}.}
\end{figure}

Here d-band of the transition metal (Cu) is split into the lower and
upper Hubbard bands by the on-site repulsive interaction $U$, while
the first band to be doped is an oxygen band within the Hubbard gap.
The oxygen band is  completely filled in parent insulators, and
doped p-holes  interact with phonons and with spin fluctuations of
d-band electrons. A characteristic magnetic interaction, which could
be  responsible for  pairing  is the spin-exchange interaction,
$J=4t^{2}/U$, of the order of $0.1$ eV (here $t$ is the hopping
integral). On the other hand, a simple parameter-free estimate of
the Fr\"ohlich electron-phonon interaction (routinely neglected
within the Hubbard $U$ and/or $t-J$ approach) yields the effective
attraction as high as $1$ eV \cite{alebook1}. This estimate  is
obtained using the familiar expression for the polaron level shift,
$E_{p},$  the high-frequency, $\epsilon _{\infty }$, and the static,
$\epsilon _{0},$ dielectric constants of the host insulator,
measured experimentally \cite{alebra1},
\begin{equation}
E_{p}={\frac{1}{{2\kappa }}}\int_{BZ}{\frac{d^{3}q}{{(2\pi )^{3}}}}{\frac{%
4\pi e^{2}}{{q^{2}}}},
\end{equation}
where $\kappa ^{-1}=\epsilon _{\infty }^{-1}-\epsilon _{0}^{-1}$ and
the size of the integration region is the Brillouin zone (BZ). Since
$\epsilon _{\infty }=5$ and $\epsilon_0=30$ in La$_2$CuO$_4$ one
obtains $E_p=0.65$ eV. Hence the attraction, which is about $2E_p$,
induced by the long-range lattice deformation in parent cuprates
 is one order of magnitude larger than the exchange
magnetic interaction. There is virtually no screening of
 e-ph interactions with $c-$axis polarized optical phonons in
doped cuprates because the upper limit for an out-of-plane plasmon
frequency ($< 200$ cm$^{-1}$)\cite{mar1} is well below
characteristic phonon frequencies, $\omega\approx$ 400 - 1000 cm
$^{-1}$ . Hence the Fr\"ohlich interaction remains the most
essential pairing interaction at any doping.

Further compelling evidence for the strong e-ph interaction has come
from isotope effects  \cite{zhao}, more recent high resolution angle
resolved photoemission spectroscopies (ARPES)
 \cite{LAN}, and a number of earlier optical \cite{mic1,ita,tal,tim} and
 neutron-scattering \cite{ega} studies of cuprates.   The strong
coupling with optical phonons, unambiguously established in all
high-temperature superconductors,  transforms holes into lattice
mobile polarons and mobile superconducting bipolarons as has been
proposed \cite{aleran,ale0}  prior the discovery \cite{mul,chu}.

When  the e-ph interaction  binds holes into intersite oxygen
bipolarons \cite{alebook1},  the chemical potential remains pinned
inside the charge transfer gap. It is found
 at a half of  the bipolaron binding energy, Fig.1, above the oxygen band edge
 shifted by the polaron level shift $E_{p}$, as clearly observed
in the tunnelling experiments by Bozovic et al. in optimally doped
La$_{1.85}$Sr$_{0.15}$CuO$_4$ \cite{boz0}. The bipolaron binding
energy as well as the singlet-triplet bipolaron exchange energy
(section 3) are thought to be the origin of normal state charge and
spin pseudogaps, respectively, as  has been proposed by us
\cite{alegap} and later found  experimentally \cite{kabmic}. In
overdoped samples carriers screen part of the e-ph interaction with
low frequency phonons. Hence, the bipolaron binding energy decreases
\cite{alekabmot} and the hole bandwidth increases with doping. As a
result, the chemical potential could enter the oxygen band in
overdoped samples  because of an overlap of the bipolaron and
polaron bands, so a Fermi-level crossing could be seen in ARPPES at
overdoping where mobile bipolarons coexist with degenerate polarons
\cite{narlikar}.

\section{The "Fr\"ohlich-Coulomb" Model (FCM)}

\subsection{Canonically transformed Hamiltonian}
Experimental facts tell us that any realistic
 description of
 high temperature superconductivity
 should treat the long-range Coulomb and
\emph{unscreened} e-ph interactions on an equal footing. In the past
decade we have developed a "Fr\"ohlich-Coulomb" model (FCM)
\cite{ale5,alebook1,alekor} to deal with the strong long-range
Coulomb and the strong long-range
 e-ph interactions in cuprates and other related compounds.
The model Hamiltonian explicitly includes a long-range
electron-phonon and the Coulomb interactions as well as the kinetic
and deformation energies.  The implicitly present large Hubbard $U$
term prohibits double occupancy and removes the need to distinguish
fermionic spins since the exchange interaction is negligible
compared with the direct Coulomb and the electron-phonon
interactions.

 Introducing  fermionic, $c_{\bf n}$, and
phononic, $d_{{\bf m}\alpha }$, operators the Hamiltonian of the
model  is written as
\begin{eqnarray}
H = & - & \sum_{\bf n \neq n'} \left[ T({\bf n-n'}) c_{\bf
n}^{\dagger } c_{\bf n'} - {1\over{2}} V_{c}({\bf n-n'}) c_{\bf
n}^{\dagger}
c_{\bf n}c_{\bf n'}^{\dagger } c_{\bf n'} \right]  \nonumber \\
& - &  \sum_{\alpha,\bf n m} \omega_{\alpha} g_{\alpha}({\bf m-n})
({\bf e}_{\alpha } \cdot {\bf u}_{\bf m-n}) c_{\bf n}^{\dagger }
c_{\bf n}
(d_{{\bf m}\alpha}^{\dagger}+d_{{\bf m}\alpha }) \nonumber \\
& + &
 \sum_{{\bf m}\alpha} \omega_{\alpha}\left( d_{{\bf m}\alpha
}^{\dagger} d_{{\bf m}\alpha }+1/2 \right),
\end{eqnarray}
where $T({\bf n})$ is the hopping integral in a rigid lattice, ${\bf
e}_{ \alpha}$ is the polarization vector of the $\alpha$th vibration
coordinate, ${\bf u}_{\bf m-n} \equiv ({\bf m-n})/|{\bf m-n}|$ is
the unit vector in the direction from electron ${\bf n}$ to  ion
${\bf m}$, $g_{\alpha}({\bf m-n)}$ is the dimensionless e-ph
coupling function, and $V_{c}({\bf n-n'})$ is the inter-site Coulomb
repulsion. $g_{\alpha}({\bf m-n)}$ is proportional to the {\em
force} $f_{\mathbf{m}}(\mathbf{n})$ acting between the electron on
site ${\bf n}$ and the ion on ${\bf m}$. For simplicity, we assume
that all the phonon modes are non-dispersive with the frequency
$\omega_{\alpha}$, and include spin in the definition of ${\bf n}$.
We also use $\hbar =k_B=c=1$.

The phonon frequency dispersion can be readily included in the
definition of all essential physical quantities such as the polaron
level shift, mass and the polaron-polaron interaction using the
quasi-momentum representation for phonons \cite{alebook1}. Generally
it leads to a lighter polaron  compared with the nondispersive
approximation. For example,  comprehensive studies of the molecular
Holstein model, in which the dispersive features of the phonon
spectrum are taken into account, found much lower values of the
polaron mass compared with the non-dispersive model \cite{zoli}.

If the electron-phonon interaction is strong, i.e. the conventional
e-ph coupling constant of the BCS theory is large, $\lambda >1$,
then the weak-coupling BCS \cite{bcs} and the intermediate-coupling
Migdal-Eliashberg \cite{mig,eli} approaches cannot be applied
\cite{alebreak}. Nevertheless the Hamiltonian, Eq.(2), can be solved
analytically  using the $"1/\lambda"$ multi-polaron expansion
technique \cite{alebook1}, if $\lambda =E_p/zT(a) >1$. Here the
polaron level shift is
\begin{equation}
 E_{p} = \sum_{{\bf n}
\alpha} \omega_{\alpha} g_{\alpha}^{2}({\bf n}) ({\bf
e}_{\alpha}\cdot {\bf u}_{\bf n})^{2} ,
\end{equation}
and $zT(a)$ is about the half-bandwidth in a rigid lattice. The
model shows a rich phase diagram depending on the ratio of the
inter-site Coulomb repulsion $V_{c}$ and the polaron level shift
$E_{p}$ \cite{alekor}. The ground state of FCM is a \emph{polaronic}
Fermi liquid when the Coulomb repulsion is large, a
\emph{bipolaronic} high-temperature superconductor at intermediate
Coulomb repulsions, and a charge-segregated insulator if the
repulsion is weak. FCM predicts \emph{superlight } polarons and
bipolarons in cuprates with a remarkably  high superconducting
critical temperature. Cuprate bipolarons are relatively light
because they are \emph{inter-site} rather than  \emph{on-site} pairs
due to the strong on-site repulsion,  and because mainly $c$-axis
polarized optical phonons are responsible for the in-plane mass
renormalization. The relatively small mass renormalization of
polaronic and bipolaronic carries in FCM has been confirmed
numerically using the exact Quantum Monte Carlo (QMC)
\cite{Korn2,jim2}, cluster diagonalization \cite{feh3} and
variational \cite{bon2} algorithms.

The $``1/\lambda $'' expansion technique is based on the fact, known
for a long time, that there is an analytical exact solution of a
$single$ polaron problem in the strong-coupling limit $\lambda
\rightarrow \infty $. Following Lang and Firsov \cite{lan} we apply
the canonical transformation $e^{S}$   diagonalising the
Hamiltonian, Eq.(2). The diagonalisation is exact, if $T({\bf m})=0$
(or $\lambda =\infty $). In the Wannier representation for electrons
and phonons,
\[
S=\sum_{{\bf m\neq n,}\alpha }g_{\alpha }({\bf m-n})({\bf e}_{\alpha }\cdot {\bf u}_{%
{\bf m-n}})c_{{\bf n}}^{\dagger }c_{{\bf n}}(d_{{\bf m}\alpha }^{\dagger }-d_{%
{\bf m}\alpha }).
\]

The transformed Hamiltonian is
\begin{eqnarray}
\tilde{H} &=&e^{-S}He^{S}=- \sum_{{\bf n\neq n^{\prime }}}\hat{\sigma}_{{\bf %
nn^{\prime }}}c_{{\bf n}}^{\dagger }c_{{\bf n^{\prime }}}+\omega\sum_{%
{\bf m}\alpha }\left( d_{{\bf m}\alpha}^{\dagger }d_{{\bf m}\alpha}+\frac{1}{2}%
\right) + \\
&&{1\over{2}}\sum_{{\bf n\neq n^{\prime }}}v({\bf n-n^{\prime
}})c_{{\bf n}}^{\dagger
}c_{{\bf n}}c_{{\bf n^{\prime }}}^{\dagger }c_{{\bf n^{\prime }}}-E_{p}\sum_{%
{\bf n}}c_{{\bf n}}^{\dagger }c_{{\bf n}} \nonumber,
\end{eqnarray}
where for simplicity we take $\omega_{\alpha}=\omega$. The last term
describes the energy gained by polarons due to the e-ph interaction.
The third term on the right-hand side  is the polaron-polaron
interaction,
\begin{equation}
v({\bf n-n^{\prime }})=V_{c}({\bf n-n^{\prime }})-V_{ph}({\bf n-n^{\prime }}%
),
\end{equation}
where
\begin{eqnarray*}
V_{ph}({\bf n-n^{\prime }}) &=&2\omega \sum_{{\bf m,}\alpha }g_{\alpha }({\bf %
m-n})g_{\alpha }({\bf m-n^{\prime }})\times \\
&&({\bf e}_{\alpha}\cdot {\bf u}_{{\bf m-n}})({\bf e}_{\alpha }\cdot {\bf u}_{%
{\bf m-n^{\prime }}}).
\end{eqnarray*}
The phonon-induced interaction $V_{ph}$ is due to displacements of
common
ions caused by two electrons. Finally, the transformed hopping operator $\hat{\sigma%
}_{{\bf nn^{\prime }}}$  is given by
\begin{eqnarray}
\hat{\sigma}_{{\bf nn^{\prime }}} &=&T({\bf n-n^{\prime }})\exp \left[ \sum_{%
{\bf m,}\alpha }\left[ g_{\alpha}({\bf m-n})({\bf e}_{\alpha }\cdot {\bf u}_{{\bf m-n%
}})\right. \right. \\
&&-\left. \left. g_{\alpha }({\bf m-n^{\prime }})({\bf e}_{\alpha }\cdot {\bf u}_{%
{\bf m-n^{\prime }}})\right] (d_{{\bf m}\alpha }^{\dagger }-d_{{\bf
m}\alpha })\right] \nonumber.
\end{eqnarray}
This term is a perturbation at  large $\lambda $.  It accounts for
the polaron and \emph{bipolaron} tunnelling and high temperature
superconductivity \cite{alebook1}. In  particular crystal structures
like perovskites, a bipolaron tunnneling could appear already in the
first order in $T({\bf n})$ (see below), so that $\hat{\sigma}_{{\bf
nn^{\prime }}}$ can be averaged over  phonon vacuum, if the
temperature is low enough, $T\ll \omega$. The result is
\begin{equation}
t({\bf n-n^{\prime }})\equiv \left\langle \left\langle \hat{\sigma}_{{\bf %
nn^{\prime }}}\right\rangle \right\rangle _{ph}=T({\bf n-n^{\prime
}})\exp [-g^{2}({\bf n-n^{\prime }})],
\end{equation}
where
\begin{eqnarray*}
g^{2}({\bf n-n^{\prime }}) &=&\sum_{{\bf m},\alpha }g_{\alpha }({\bf m-n})({\bf e}%
_{\alpha }\cdot {\bf u}_{{\bf m-n}})\times \\
&&\left[ g_{\alpha}({\bf m-n})({\bf e}_{\alpha }\cdot {\bf u}_{{\bf
m-n}})-g_{\alpha
}({\bf m-n^{\prime }})({\bf e}_{\alpha }\cdot {\bf u}_{{\bf m-n^{\prime }}})%
\right] .
\end{eqnarray*}
By comparing Eqs.(7) and Eqs.(3,5), the bandwidth renormalization
exponent can be expressed via $E_{p}$ and $V_{ph}$ as follows
\begin{equation}
g^{2}({\bf n-n^{\prime }})=\frac{1}{\omega}\left[ E_{p}-\frac{1}{2}%
V_{ph}({\bf n-n^{\prime }})\right] .
\end{equation}
In zero order with respect to the hopping the Hamiltonian, Eq.(4)
describes localised polarons and independent phonons, which are
vibrations of ions around new equilibrium positions depending on the
polaron occupation numbers. The phonon frequencies remain unchanged
in this limit. The middle of the electron band falls by the polaron
level-shift $E_{p}$ due to a potential well created by lattice
deformation. The finite hopping term leads to the polaron tunnelling
because of degeneracy of the zero order Hamiltonian with respect to
 site positions of the polaron.

\subsection{Superlight small bipolarons in FCM: root to room temperature superconductivity}

Now let us consider  in-plane bipolarons in a two-dimensional
lattice of ideal octahedra that can be regarded as a simplified
model of the copper-oxygen perovskite layer, Fig.2 \cite{alekor}.
The lattice period is $a=1$ and the distance between the apical
sites and the central plane is $h=a/2=0.5$. For mathematical
transperancy we assume that all in-plane atoms, both copper and
oxygen, are static but apex oxygens are independent
three-dimensional isotropic harmonic oscillators.

\begin{figure}[tbp]
\begin{center}
\includegraphics[angle=-0,width=0.47\textwidth]{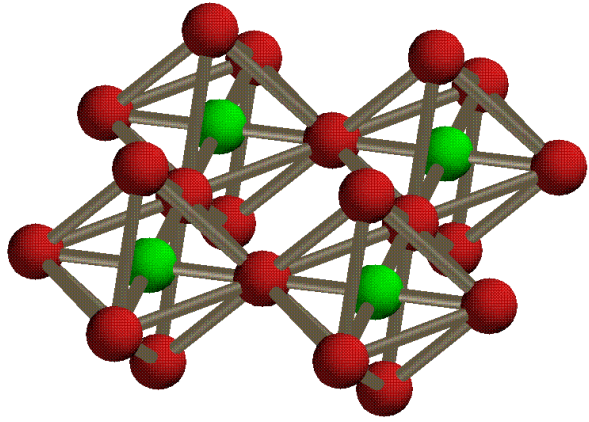} \vskip -0.5mm
\end{center}
\caption{Simplified model of the copper-oxygen perovskite layer
\cite{alekor}. }
\end{figure}
 Due to poor
screening, the hole-apex interaction is purely coulombic,
\[
g_{\alpha }({\bf m-n})=\frac{\kappa _{\alpha }}{|{\bf m-n}|^{2}},
\]
where $\alpha =x,y,z$. To account for the experimental fact that
$z$-polarized
phonons couple to the holes stronger than others \cite{tim}, we choose $%
\kappa _{x}=\kappa _{y}=\kappa _{z}/\sqrt{2}$. The direct hole-hole
repulsion is
\[
V_{c}({\bf n-n^{\prime }})=\frac{V_{c}}{\sqrt{2}|{\bf n-n^{\prime
}}|}
\]
so that the repulsion between two holes in the nearest neighbour
(NN) configuration is $V_{c}$. We also include the bare NN hopping
$T_{NN}$, the next nearest neighbour (NNN) hopping across copper
$T_{NNN}$ and the NNN hopping between the pyramids $T_{NNN}^{\prime
}$.

\begin{figure}[tbp]
\begin{center}
\includegraphics[angle=-0,width=0.50\textwidth]{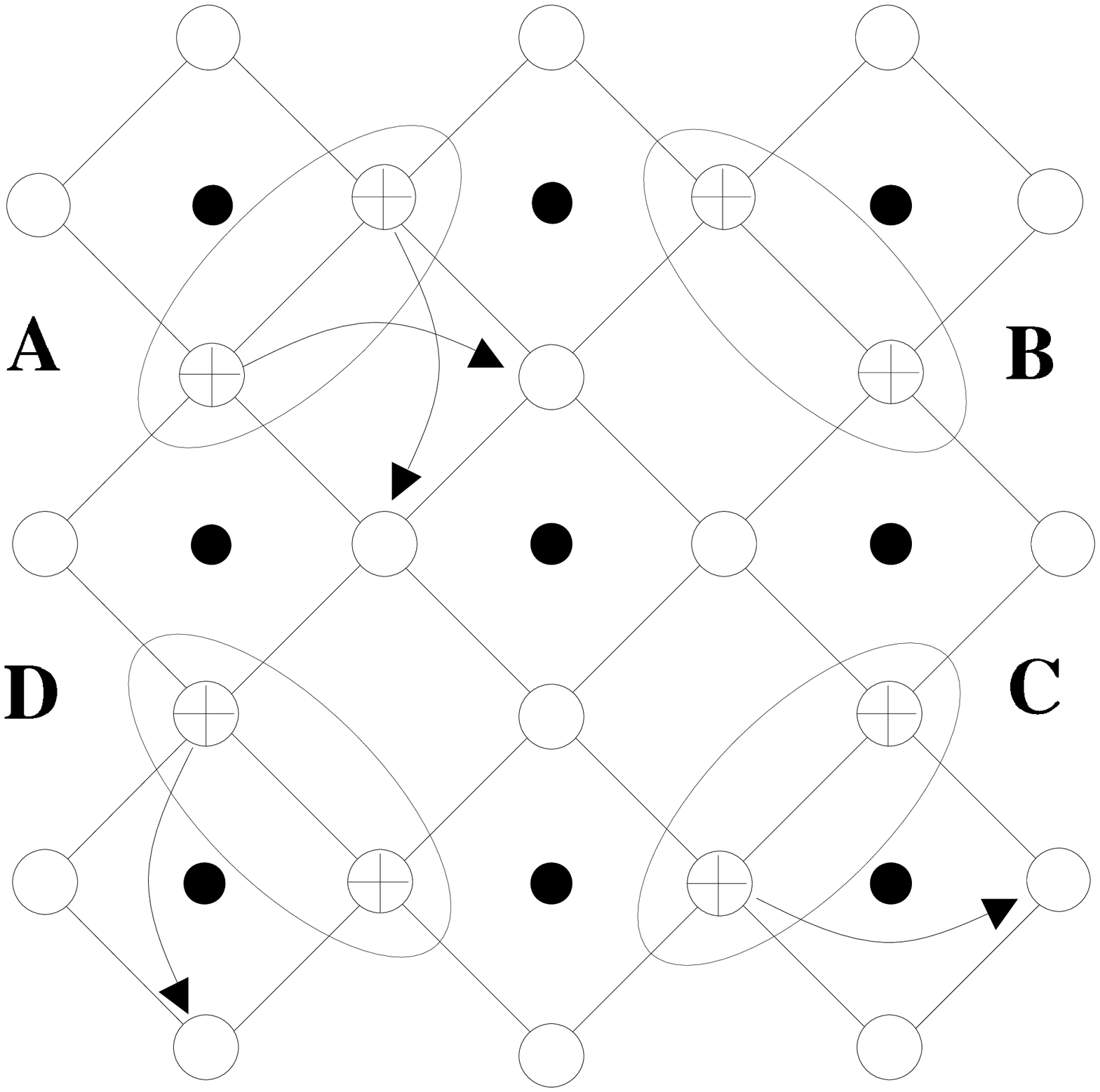} \vskip -0.5mm
\end{center}
\caption{Four degenerate in-plane bipolaron configurations A, B, C,
and D . Some single-polaron hoppings are indicated by arrows
\cite{alekor}. }
\end{figure}

The polaron shift is given by the lattice sum Eq.(3), which after
summation over polarizations yields
\begin{equation}
E_{p}=2\kappa _{x}^{2}\omega _{0}\sum_{{\bf m}}\left( \frac{1}{|{\bf m-n}%
|^{4}}+\frac{h^{2}}{|{\bf m-n}|^{6}}\right) =31.15\kappa
_{x}^{2}\omega _{0}, \label{televen}
\end{equation}
where the factor $2$ accounts for two layers of apical sites. For
reference,
the Cartesian coordinates are ${\bf n}=(n_{x}+1/2,n_{y}+1/2,0)$, ${\bf m}%
=(m_{x},m_{y},h)$, and $n_{x},n_{y},m_{x},m_{y}$ are integers. The
polaron-polaron attraction is
\begin{equation}
V_{ph}({\bf n-n^{\prime }})=4\omega \kappa _{x}^{2}\sum_{{\bf m}}\frac{%
h^{2}+({\bf m-n^{\prime }})\cdot ({\bf m-n})}{|{\bf m-n^{\prime }}|^{3}|{\bf %
m-n}|^{3}}.  \label{ttwelve}
\end{equation}
Performing the lattice summations for the NN, NNN, and NNN'
configurations one finds $V_{ph}=1.23\,E_{p},$ $0.80\,E_{p}$, and
$0.82\,E_{p}$,
respectively. As a result, we obtain a net inter-polaron interaction as $%
v_{NN}=V_{c}-1.23\,E_{p}$, $v_{NNN}=\frac{V_{c}}{\sqrt{2}}-0.80\,E_{p}$, $%
v_{NNN}^{\prime }=\frac{V_{c}}{\sqrt{2}}-0.82\,E_{p}$, and the mass
renormalization exponents as $g_{NN}^{2}=0.38(E_{p}/\omega)$, $%
g_{NNN}^{2}=0.60(E_{p}/\omega)$ and $(g^{\prime
}{}_{NNN})^{2}=0.59(E_{p}/\omega)$.

Let us now discuss different regimes of the model. At
$V_{c}>1.23\,E_{p}$, no bipolarons are formed and the system is a
polaronic Fermi liquid. Polarons tunnel in the {\em square} lattice
with  $t=T_{NN}\exp (-0.38E_{p}/\omega)$ and  $t^{\prime
}=T_{NNN}\exp (-0.60E_{p}/\omega)$ for NN and NNN hoppings,
respectively. Since $g_{NNN}^{2}\approx (g_{NNN}^{\prime })^{2} $
one can neglect the difference between NNN hoppings within and
between the octahedra. A single polaron spectrum is therefore
\begin{equation}
E_{1}({\bf k})=-E_{p}-2t^{\prime }[\cos k_{x}+\cos k_{y}]-4t\cos
(k_{x}/2)\cos (k_{y}/2).  \label{tfifteen}
\end{equation}
The polaron mass is $m^{\ast }=1/(t+2t^{\prime })$. Since in general $%
t>t^{\prime }$, the mass is mostly determined by the NN hopping amplitude $t$%
.

If $V_{c}<1.23\,E_{p}$ then intersite NN bipolarons form. The
bipolarons
tunnel in the plane via four resonating (degenerate) configurations $A$, $B$%
, $C$, and $D$, as shown in Fig.3. In the first order of the
renormalised hopping integral, one should retain only these lowest
energy configurations and discard all the processes that involve
configurations with higher energies. The result of such a projection
is the bipolaron Hamiltonian
\begin{eqnarray}
H_{b} &=&(V_{c}-3.23\,E_{p})\sum_{{\bf l}}[A_{{\bf l}}^{\dagger }A_{{\bf l}%
}+B_{{\bf l}}^{\dagger }B_{{\bf l}}+C_{{\bf l}}^{\dagger }C_{{\bf l}}+D_{%
{\bf l}}^{\dagger }D_{{\bf l}}] \\
&&-t^{\prime }\sum_{{\bf l}}[A_{{\bf l}}^{\dagger }B_{{\bf l}}+B_{{\bf l}%
}^{\dagger }C_{{\bf l}}+C_{{\bf l}}^{\dagger }D_{{\bf l}}+D_{{\bf l}%
}^{\dagger }A_{{\bf l}}+H.c.]  \nonumber \\
&&-t^{\prime }\sum_{{\bf n}}[A_{{\bf l-x}}^{\dagger }B_{{\bf l}}+B_{{\bf l+y}%
}^{\dagger }C_{{\bf l}}+C_{{\bf l+x}}^{\dagger }D_{{\bf l}}+D_{{\bf l-y}%
}^{\dagger }A_{{\bf l}}+H.c.],  \nonumber
\end{eqnarray}
where ${\bf l}$ numbers octahedra rather than individual sites, ${\bf x}%
=(1,0)$, and ${\bf y}=(0,1)$. A Fourier transformation and
diagonalization of a $4\times 4$ matrix yields the bipolaron
spectrum:
\begin{equation}
E_{2}({\bf K})=V_{c}-3.23E_{p}\pm 2t^{\prime }[\cos (K_{x}/2)\pm
\cos (K_{y}/2)].  \label{tseventeen}
\end{equation}
There are four bipolaronic subbands combined in the band of the width $%
8t^{\prime }$. The effective mass of the lowest band is $m^{\ast
\ast }=2/t^{\prime }$. The bipolaron binding energy is $\Delta
\approx 1.23E_{p}-V_{c}.$ Inter-site bipolarons already move in the
{\em first} order of the single polaron hopping. This remarkable
property is entirely due to the strong on-site repulsion and
long-range electron-phonon interactions that leads to a non-trivial
connectivity of the lattice. This fact combines with a weak
renormalization of $t^{\prime }$ yielding a {\em superlight}
bipolaron with the mass $m^{\ast \ast }\propto \exp
(0.60\,E_{p}/\omega )$. We recall that in the Holstein model
$m^{\ast \ast }\propto \exp (2E_{p}/\omega )$ \cite{aleran}. Thus
the mass of the Fr\"{o}hlich bipolaron in the perovskite layer
scales approximately as a {\em cubic root} of that of the Holstein
bipolaron.

At even stronger e-ph interaction, $V_{c}<1.16E_{p}$, NNN bipolarons
become stable. More importantly, holes can now form 3- and
4-particle clusters. The dominance of the potential energy over
kinetic in the transformed Hamiltonian enables us to readily
investigate these many-polaron cases. Three holes placed within one
oxygen square have four degenerate states with the energy
$2(V_{c}-1.23E_{p})+V_{c}/\sqrt{2}-0.80E_{p}$. The first-order
polaron hopping processes mix the states resulting in a ground state
linear
combination with the energy $E_{3}=2.71V_{c}-3.26E_{p}-\sqrt{%
4t^{2}+t^{\prime }{}^{2}}$. It is essential that between the squares
such triads could move only in higher orders of polaron hopping. In
the first order, they are immobile. A cluster of four holes has only
one state within
a square of oxygen atoms. Its energy is $E_{4}=4(V_{c}-1.23E_{p})+2(V_{c}/%
\sqrt{2}-0.80E_{p})=5.41V_{c}-6.52E_{p}$. This cluster, as well as
all bigger ones, are also immobile in the first order of polaron
hopping. We would like to stress that at distances much larger than
the lattice constant the polaron-polaron interaction is always
repulsive, and the formation of infinite clusters, stripes or
strings is  prohibited \cite{alekabstring}. Hence the long-range
Fr\"ohlich interaction combined with Coulomb repulsion might cause
clustering of polarons into finite-size quasi-metallic mesoscopic
textures. However analytical \cite{alekabstring} and QMC \cite{kab}
studies of mesoscopic textures with lattice deformations and Coulomb
repulsion show that pairs (i.e. bipolarons) dominate over phase
separation since they effectively repel each other \cite{alebook1}.

As shown above  FCM is reduced to an extended Hubbard model with
intersite attraction and suppressed double-occupancy in the limit of
high phonon frequency $\omega\gtrsim T(a) $ and large on-site
Coulomb repulsion. Then the Hamiltonian can be projected onto the
subspace of nearest neighbor intersite  bipolarons, Fig.3. In
contrast with the crawler motion of on-site bipolaron, the intersite
bipolaron tunnelling is a crab-like, so that  its mass scales
linearly with the polaron mass ($m^{**}\approx 4m^*$ on the
staggered chain \cite{alekor}). To study FCM for more realistic
intermediate values of the electron-phonon coupling and phonon
frequency  the CTQMC algorithm  \cite{kor,Korn2} has been recently
extended to systems of two particles with strong electron-phonon
interactions. We have solved the bipolaron problem on a staggered
ladder, triangular and anisotropic hexagonal lattices from weak to
strong coupling \cite{jim2} in a realistic parameter range where
usual limiting approximations fail. The bipolaron to polaron mass
ratio has been found about 2 in the weak coupling regime
($\lambda\ll1$) as it should be for a large bipolaron \cite{dev}. In
the strong-coupling, large phonon frequency limit the mass ratio
approaches 4, in agreement with strong-coupling arguments given
above.  In a wide region of parameter space, we find a
bipolaron/polaron mass ratio of between 2 and 4 and a bipolaron
radius similar to the lattice spacing. Thus the bipolaron is small
and light at the same time. Taking into account additional intersite
Coulomb repulsion $V_c$ does not change this conclusion.  As $V_c$
increases the bipolaron mass decreases but the radius remains small,
at about 2 lattice spacings.

When bipolarons are small and pairs do not overlap, the pairs can
form  a Bose-Einstein condensate (BEC). Our CTQMC simulations show
that with realistic values for the coupling constant, $\lambda\simeq
1$, and phonon frequencies, $\omega \simeq T(a)$ one can avoid
overlap of pairs and get the bose-condensation temperature $T_{c}$
about the room temperature. We believe that the following recipe is
worth investigating to look for room-temperature superconductivity
\cite{jim2}: (a) The parent compound should be an ionic insulator
with light ions to form high-frequency optical phonons, (b) The
structure should be quasi two-dimensional to ensure poor screening
of high-frequency c-axis polarized phonons, (c) A triangular lattice
is required in combination with  strong, on-site
 Coulomb repulsion to form the small superlight Crab
bipolaron  (d) Moderate carrier densities are required to keep the
system of small bipolarons close to the dilute regime.

 There are strong
arguments  in favor of 3D bipolaronic BEC in cuprates
\cite{alebook1} drawn using parameter-free fitting of experimental
$T_c$ with BEC $T_{c}$ \cite{alekab}, unusual upper critical fields
\cite{aleH} and the specific heat \cite{alekablia}, and, more
recently normal state  diamagnetism \cite{aledia},  the Hall-Lorenz
numbers \cite{leeale,alelor}, the normal state Nernst effect
\cite{alezav,alecon},
 and the giant proximity effect (GPE)
\cite{alepro} as discussed below.

\section{Some normal state properties of cuprates in FCM}
\begin{figure}
\begin{center}
\includegraphics[angle=-90,width=0.80\textwidth]{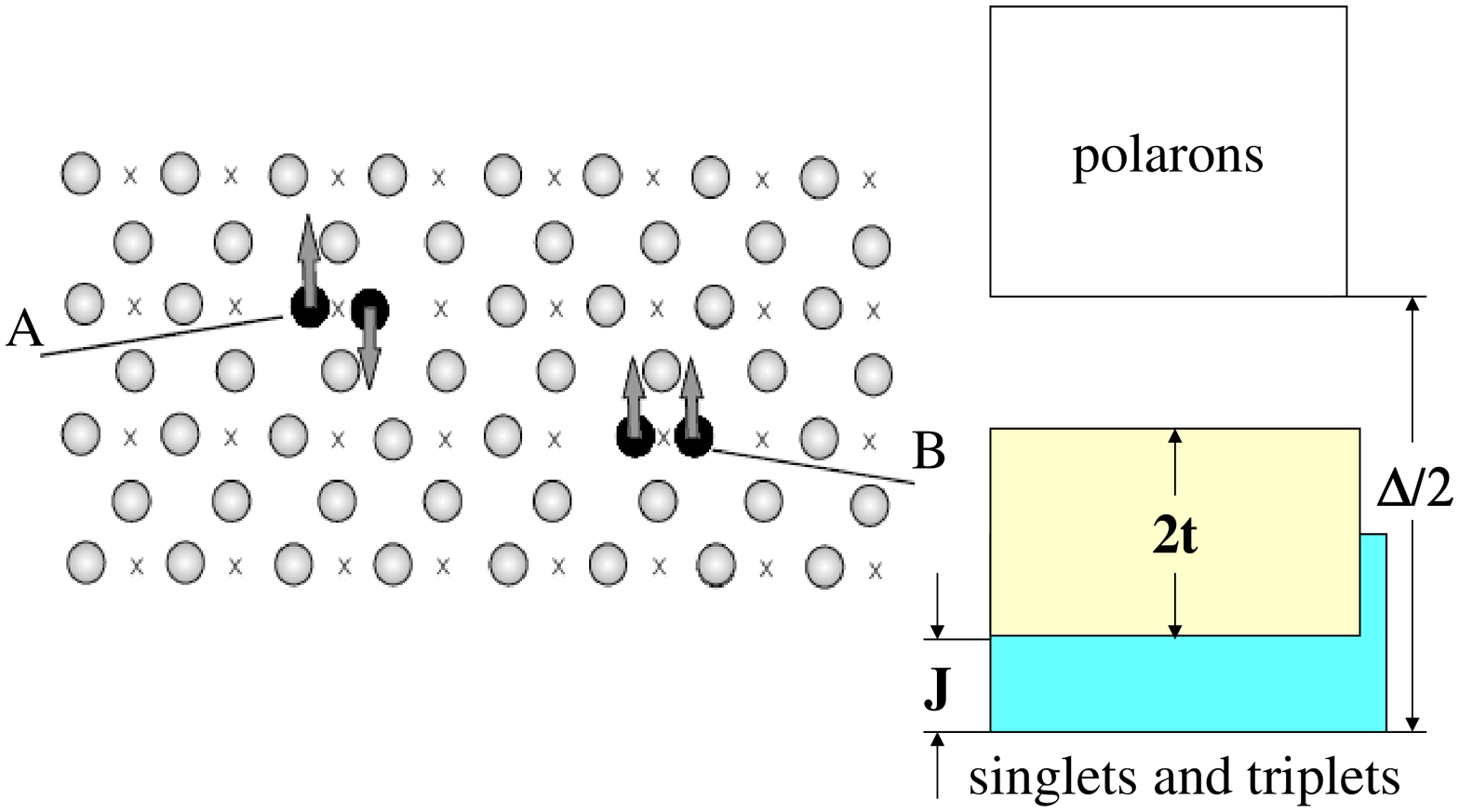}
\vskip -0.5mm \caption{Bipolaron picture of high temperature
superconductors. $A$ corresponds to a singlet oxygen intersite
bipolaron, $B$ is a triplet intersite bipolaron, $\Delta$ is the
singlet bipolaron binding energy, $J$ is the singlet-triplet
exchange energy, and $2t$ is the bipolaron bandwidth
\cite{alemot3}.}
\end{center}
\end{figure}

\subsection{Normal state Nernst effect and insulating-like in-plane resistivity}

The low-energy FCM electronic structure  of  cuprates  is shown in
Fig.4 \cite{alemot3}. Polaronic p-holes  are bound in lattice
inter-site singlets (A) or in singlets and  triplets (B) (if spins
are included in Eq.(2)) at any temperature. Above T$_{c}$ a charged
bipolaronic  Bose liquid is non-degenerate and below $T_{c}$ phase
coherence (ODLRO) of the preformed bosons sets in. The state above
$T_{c}$ is perfectly "normal" in the sense that the off-diagonal
order parameter (i.e. the Bogoliubov-Gor'kov anomalous average
$\cal{F}(\mathbf{r,r^{\prime }})=\langle \psi_{\downarrow
}(\mathbf{{r})\psi _{\uparrow }({r^{\prime }}\rangle}$) is zero
above the resistive transition temperature $T_{c}$. Here
$\psi_{\downarrow,\uparrow }(\mathbf{r})$ annihilates  electrons
with spin $\downarrow, \uparrow$ at point ${\bf r}$.

In contrast with our bipolaronic (and with BCS) theory a significant
fraction of research in the field of cuprate superconductors
suggests a so-called phase fluctuation scenario, where
$\cal{F}(\mathbf{r,r^{\prime }})$ remains nonzero well above
$T_{c}$. I believe that the phase fluctuation scenario  is
impossible to reconcile with the extremely sharp resistive
transitions at $T_c$ in high-quality underdoped, optimally doped and
overdoped cuprates. For example, the in-plane and out-of-plane
resistivity of $Bi-2212$, where the anomalous Nernst signal has been
measured \cite{xu}, is perfectly "normal" above $T_c$, Fig.5,
showing only a few percent positive or negative magnetoresistance
\cite{zavale}, explained with bipolarons \cite{zavalemos}.
\begin{figure}
\begin{center}
\includegraphics[angle=-0,width=0.75\textwidth]{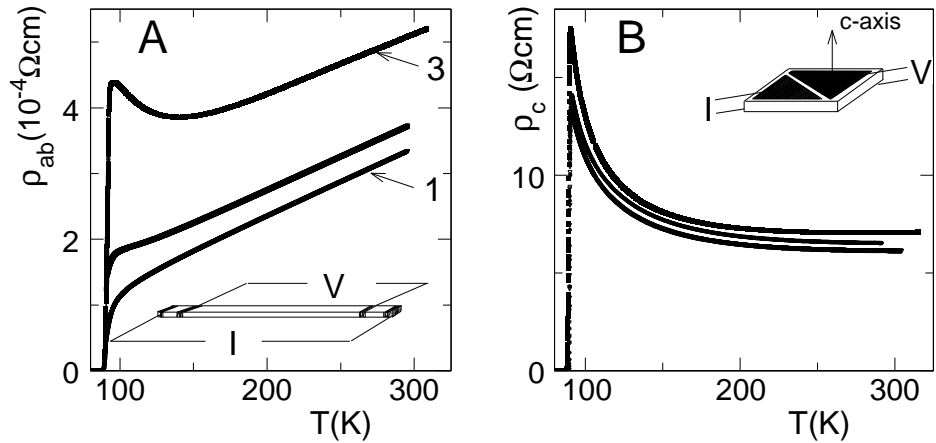}
\vskip -0.5mm \caption{In-plane (A) and out-of-plane (B) resistivity
of 3   single crystals of Bi$_2$Sr$_2$CaCu$_2$O$_8$ \cite{zavale}
showing no signature of phase fluctuations  above the resistive
transition. }
\end{center}
\end{figure}
Both in-plane \cite{buc,mac0,boz,lawrie,gan} and out-of-plane \cite
{alezavnev,hof2,zve} resistive transitions  of high-quality samples
 remain sharp in the magnetic field providing a
reliable determination of the genuine $H_{c2}(T)$.  The preformed
Cooper-pair (or phase fluctuation) model \cite{kiv}  is incompatible
with a great number of thermodynamic, magnetic, and kinetic
measurements, which show that only holes (density $x$), doped into a
parent insulator are carriers \emph{both} in  the normal and the
superconducting states of cuprates. The assumption \cite{kiv} that
the superfluid density $x$ is small compared with the normal-state
carrier density is also inconsistent with the theorem \cite{pop},
which proves that the number of supercarriers at $T=0$K  should be
the same as the number of normal-state carriers in any  clean
superfluid.

Recently we have described a number of  unusual normal state
properties in cuprates in a different manner as perfectly normal
state phenomena. In particular, the bipolaron theory  accounts for
the anomalously large  Nernst signal, the thermopower and the
insulating-like in-plane low temperature resistance
\cite{alezav,alecon} as observed \cite{xu,cap,cap2,ong}.

Thermomagnetic effects appear in conductors subjected to a
longitudinal temperature gradient $\nabla _{x}T$ in $x$ direction
and a perpendicular magnetic field  in $z$ direction. The transverse
Nernst-Ettingshausen effect \cite{nernst}  (here the Nernst effect)
is the appearance of a transverse electric field $E_y$ in the third
direction. When bipolarons are formed in the strong-coupling regime,
the chemical potential is  negative. It is found in the impurity
band just below the mobility edge at $T>T_c$. Carriers, localised
below the mobility edge contribute to the longitudinal transport
together with the itinerant carriers in extended states above the
mobility edge. Importantly the contribution of localised carriers of
any statistics to the \emph{ transverse} transport is normally small
\cite{ell} since a microscopic Hall voltage will only develop at
junctions in the intersections of the percolation paths, and it is
expected that these are few for the case of hopping conduction among
disorder-localised states \cite{mott2}. Even if this contribution is
not negligible, it adds  to the contribution of itinerant carriers
to produce a large Nernst signal, $e_{y}(T,B)\equiv -E_{y}/\nabla
_{x}T$, while it reduces the thermopower $S$ and the Hall angle
$\Theta$. This unusual "symmetry breaking" is completely at variance
with  ordinary metals where the familiar "Sondheimer" cancelation
\cite{sond} makes  $e_{y}$ much smaller than $S\tan \Theta$ because
of the electron-hole symmetry near the Fermi level. Such  behaviour
originates in the "sign" (or "$p-n$") anomaly of the Hall
conductivity of localised carriers. The sign of their Hall effect is
often $opposite$ to that of the thermopower as observed in many
amorphous semiconductors \cite{ell} and described theoretically
\cite{fri}.

The Nernst signal is expressed in terms of the kinetic coefficients
$\sigma _{ij}$ and $\alpha _{ij}$ as
\begin{equation}
e_{y}={\frac{{\sigma _{xx}\alpha _{yx}-\sigma _{yx}\alpha
_{xx}}}{{\sigma _{xx}^{2}+\sigma _{xy}^{2}}}},
\end{equation}
where the current density  is given by $j_{i}=\sigma
_{ij}E_{j}+\alpha _{ij}\nabla _{j}T$.
 When the chemical potential $\mu$ is at the mobility edge,
 localised carriers contribute to the transport,
 so  $\sigma _{ij}$ and $\alpha _{ij}$  can be expressed
as $\sigma^{ext} _{ij}+\sigma^{l}_{ij}$ and $\alpha^{ext}
_{ij}+\alpha^{l}{ij}$, respectively. Since the Hall mobility of
carriers localised below $\mu$, $\sigma^{l}_{yx}$, has the  sign
opposite to that of carries in the extended states above $\mu$,
$\sigma^{ext}_{yx}$, the sign of the off-diagonal Peltier
conductivity $\alpha^{l}_{yx}$ should be the same as the sign of
$\alpha^{ext}_{yx}$. Then  neglecting the magneto-orbital effects in
the resistivity (since $\Theta \ll 1$ \cite{xu}) we obtain
\begin{equation}
S\tan \Theta \equiv {\sigma _{yx}\alpha _{xx}\over{\sigma
_{xx}^{2}+\sigma _{xy}^{2}}} \approx\rho (\alpha ^{ext}_{xx}-|\alpha
^{l}_{xx}|) (\Theta^{ext}-|\Theta^{l}|)
\end{equation}
and
\begin{equation}
e_{y}\approx\rho (\alpha^{ext} _{yx}+|\alpha^{l} _{yx}|)-S\tan
\Theta,
\end{equation}
where $\Theta^{ext}\equiv \sigma^{ext}_{yx}/\sigma_{xx}$,
$\Theta^{l}\equiv \sigma^{l}_{yx}/\sigma_{xx}$, and
$\rho=1/\sigma_{xx}$ is the resistivity.

\begin{figure}
\begin{center}
\includegraphics[width=0.5\textwidth]{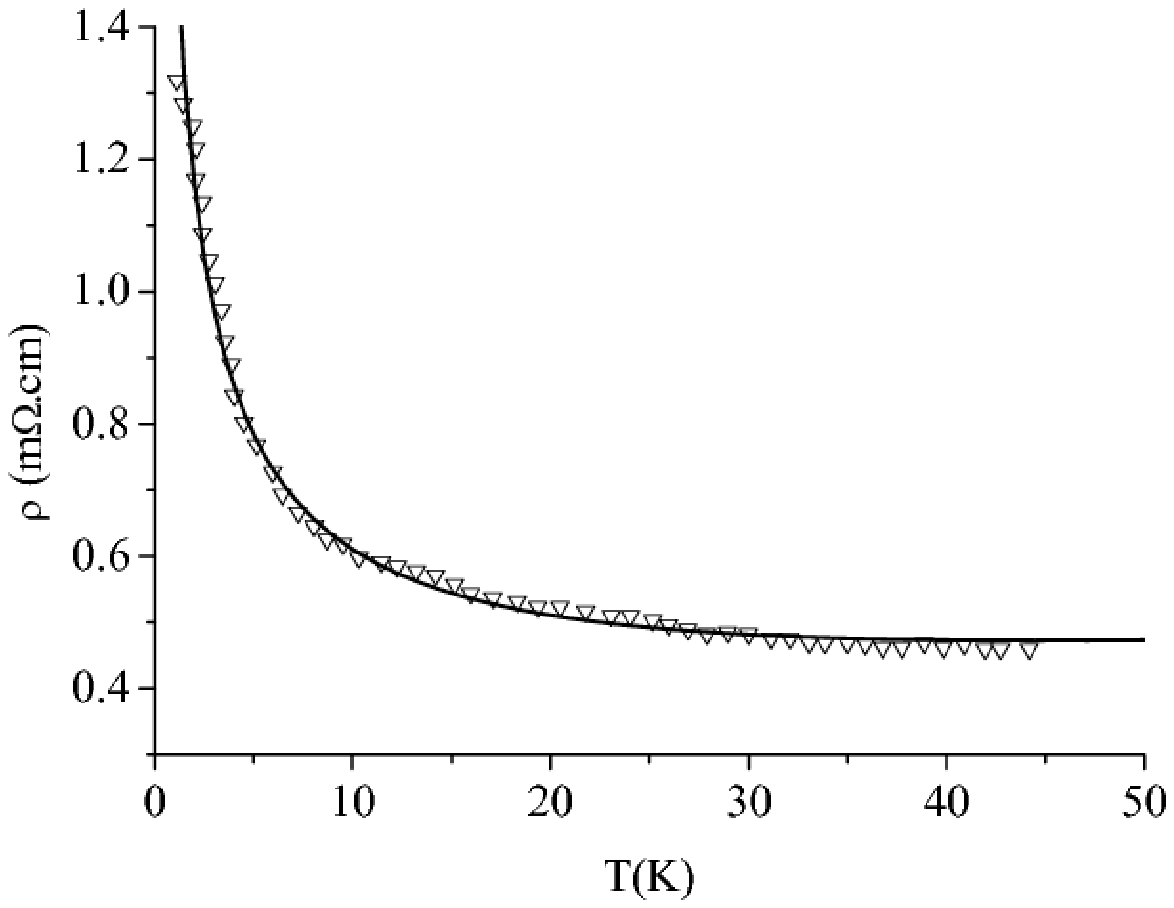}
\caption{Normal state in-plane resistivity of underdoped La$_{1.94}$
Sr$_{0.06}$CuO$_4$ (triangles \cite{cap}) as revealed in the field
$B=12$ Tesla  and compared with the bipolaron theory, Eq.(19) (solid
line).}
\end{center}
\end{figure}
Clearly the model, Eqs.(15,16) can account for a low value of
$S\tan\Theta $ compared with a large value of $e_y$ in some
underdoped cuprates \cite{xu,cap2} due to the sign anomaly. Even in
the case when  localised bosons contribute little to the
conductivity   their contribution to the thermopower $S=\rho (\alpha
^{ext}_{xx}-|\alpha ^{l}_{xx}|))$ could almost cancel  the opposite
sign contribution of itinerant carriers \cite{alezav}. Indeed the
longitudinal conductivity of itinerant two-dimensional bosons,
$\sigma^{ext} \propto \int_0 dE E df(E)/dE$ diverges logarithmically
when $\mu$ in the Bose-Einstein distribution function
$f(E)=[\exp((E-\mu)/T)-1]^{-1}$ goes to zero and the relaxation time
$\tau$ is a constant. At the same time $\alpha^{ext}_{xx}\propto
\int_0 dE E(E-\mu) df(E)/dE$ remains finite, and it could have the
magnitude comparable   with  $\alpha^{l}_{xx}$. Statistics of
bipolarons  gradually changes from Bose to Fermi statistics with
lowering energy across the mobility edge because of the Coulomb
repulsion of bosons in localised states  \cite{alegile}. Hence one
can use the same expansion near the mobility edge as in  ordinary
amorphous semiconductors to obtain the familiar textbook result
$S=S_0T$ with a constant $S_0$ at low temperatures \cite{mott3}. The
model becomes particularly simple, if we   neglect the localised
carrier contribution to $\rho$, $\Theta$ and $\alpha_{xy}$, and take
into account that $\alpha^{ext}_{xy} \propto B/\rho^2$ and
$\Theta^{ext}\propto B/\rho$ in the Boltzmann theory. Then
Eqs.(15,16) yield
\begin{equation}
S\tan \Theta  \propto T/\rho
\end{equation}
and
\begin{equation}
e_{y}(T,B)\propto (1-T/T_1)/\rho.
\end{equation}
According to our earlier suggestion \cite{alelog} the
insulating-like low-temperature dependence of $\rho(T)$ in
underdoped cuprates  originates from the elastic scattering of
nondegenerate itinerant carriers off charged  impurities. We assume
here that the carrier density is temperature independent at very low
temperatures. The relaxation time of nondegenerate carriers depends
on temperature as $\tau \propto T^{-1/2}$ for scattering off
short-range deep potential wells, and as $T^{1/2}$ for very shallow
wells \cite{alelog}. Combining both scattering rates we obtain
\begin{equation}
\rho =\rho_0[(T/T_2)^{1/2}+(T_2/T)^{1/2}].
\end{equation}
Eq.(19) with $\rho_0=0.236$ m$\Omega\cdot$cm and $T_2=44.6$K fits
extremely well the experimental insulating-like normal state
resistivity of underdoped La$_{1.94}$ Sr$_{0.06}$CuO$_4$ in the
whole low-temperature range from  2K up to 50K, Fig.6,  as revealed
in the field $B=12$ Tesla \cite{cap,cap2}. Another high quality fit
can be  obtained combining the Brooks-Herring formula for the 3D
scattering off screened charged impurities, as proposed in
Ref.\cite{kast} for almost undoped $LSCO$, or the Coulomb scattering
in 2D ($\tau \propto T$) and a temperature independent scattering
rate off neutral impurities with the carrier exchange \cite{erg}
similar to the scattering of slow electrons by hydrogen atoms in
three dimensions. Hence the scale $T_2$, which determines the
crossover toward an insulating behavior, depends on the relative
strength of two scattering mechanisms. Importantly our expressions
(17,18) for  $S\tan \Theta$ and $e_y$ do not depend on the
particular scattering mechanism. Taking into account the excellent
fit of Eq.(19) to the experiment, they can be parameterized as
\begin{equation} S\tan \Theta = e_0
{(T/T_2)^{3/2}\over{1+T/T_2}},
\end{equation}
and
\begin{equation}
e_{y}(T,B)=e_0{(T_1-T) (T/T_2)^{1/2}\over{T_2+T}} ,
\end{equation}
where $T_1$ and $e_0$ are temperature independent.
\begin{figure}
\begin{center}
\includegraphics[angle=270,width=0.70\textwidth]{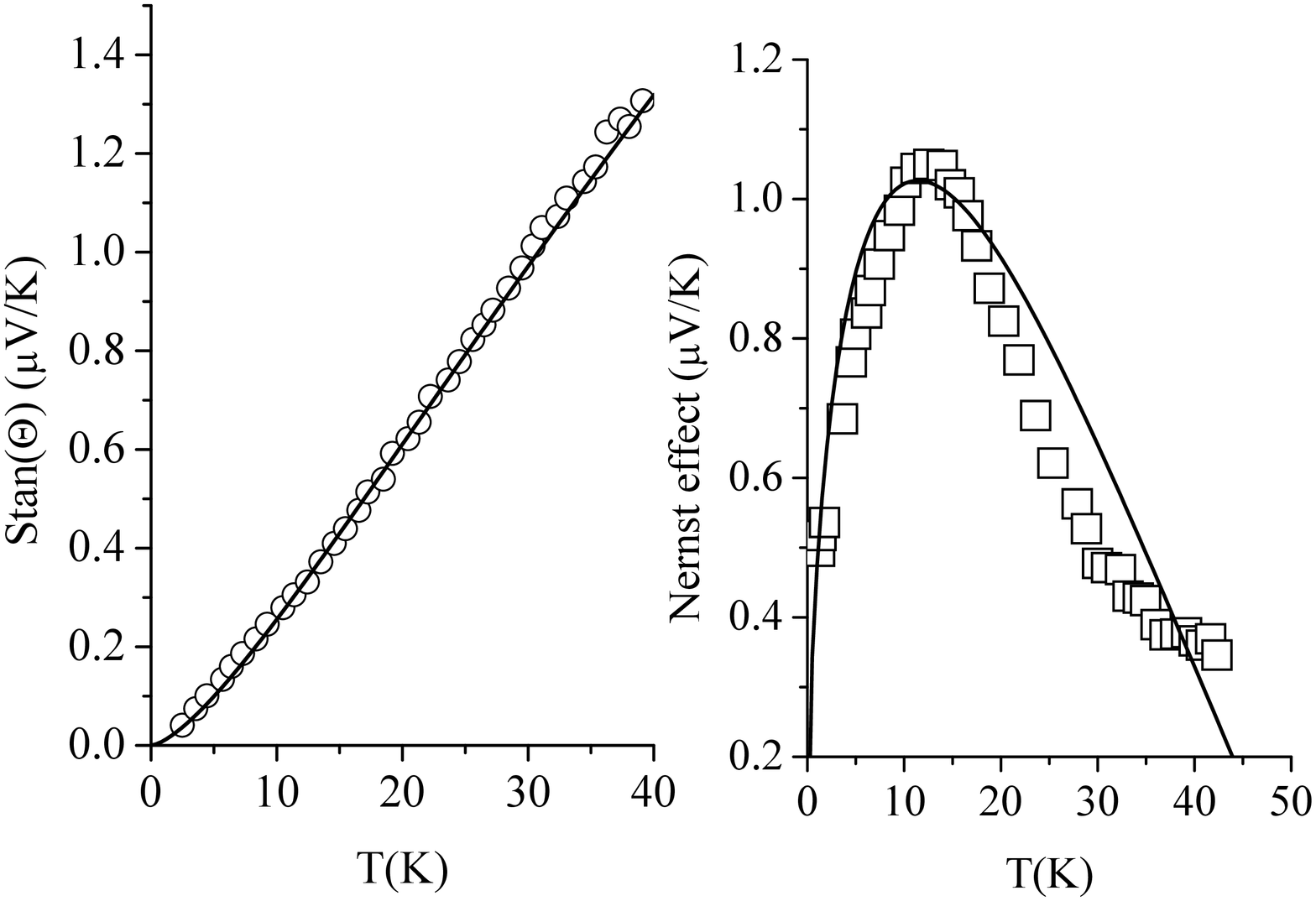}
\vskip -0.5mm \caption{$S\tan\Theta$ (circles \cite{cap2} )  and the
Nernst effect $e_y$  (squares \cite{cap})  of underdoped La$_{1.94}$
Sr$_{0.06}$CuO$_4$ at $B=12$ Tesla compared with the bipolaron
theory, Eqs.(20,21) (solid lines) \cite{alecon}.}
\end{center}
\end{figure}

In spite of all simplifications, the model describes  remarkably
well both $S\tan \Theta$ and $e_y$, Fig.7,  measured in La$_{1.94}$
Sr$_{0.06}$CuO$_4$ with a $single$ fitting parameter, $T_1=50$K
using the experimental $\rho(T)$. The constant  $e_0=2.95$ $\mu$V/K
scales the magnitudes of $S\tan \Theta$ and $e_y$.  The magnetic
field $B=12$ Tesla destroys the superconducting state of the
low-doped La$_{1.94}$ Sr$_{0.06}$CuO$_4$ down to $2$K, Fig.6, so any
residual superconducting order above $2$K is clearly ruled out,
while the Nernst signal, Fig.7, is remarkably large. The coexistence
of the large Nernst signal and a nonmetallic resistivity is in sharp
disagreement with the vortex scenario, but in agreement with our
model. Taking into account the field dependence of the conductivity
of localised carriers, the phonon-drug effect, and their
contribution to the transverse magnetotransport  can well describe
the magnetic field dependence of the Nernst signal \cite{alezav} and
improve the fit in Fig.7 at the expense of the increasing number of
fitting parameters.

\subsection{Hall-Lorenz number}

 Recent  measurements of the Righi-Leduc effect provides further evidence for  real-space charged bosons preformed above $T_c$ \cite{leeale,alelor}.
  The
effect describes transverse heat flow resulting from a perpendicular
temperature gradient in an external magnetic field, which is  a
thermal analog of the Hall effect. Using the effect the
"Hall-Lorenz" electronic number, $ L_{H}=\left( e/k_{B}\right)
^{2}\kappa _{xy}/(T\sigma _{xy})$ has been directly measured
 \cite{ZHANG} in $YBa_{2}Cu_{3}O_{6.95}$ and $YBa_{2}Cu_{3}O_{6.6}$
since transverse thermal $%
\kappa _{xy}$ and electrical $\sigma _{xy}$ conductivities involve
presumably  only  electrons. The experimental $L_{H}(T)$ showed a
quasi-linear temperature dependence above the resistive $T_{c}$,
which strongly violates the WF law. Remarkably, the measured value
of $L_{H}$ just above $T_{c}$ turned out precisely the same as
predicted by the bipolaron theory \cite{NEV}, $L=0.15L_{0}$, where
$L_{0}=\pi^2/3$ is the conventional Sommerfeld value. The breakdown
of the WF law  revealed in the Righi-Leduc effect \cite{ZHANG}  has
been explained by a temperature-dependent contribution of thermally
excited single polarons to the transverse magneto-transport
\cite{leeale}.

 Surprisingly more recent measurements of the Hall-Lorenz number
in single crystals  of optimally doped $YBa_{2}Cu_{3}O_{6.95}$ and
optimally doped and underdoped $EuBa_{2}Cu_{3}O_{y}$ led to an
opposite conclusion \cite{mat}. The experimental $L_H$ for  these
samples has turned out only weakly temperature dependent and
exceeding the Sommerfeld value by more than 2 times in the whole
temperature range from $T_c$ up to the room temperature.  Following
an earlier claim\cite{li} Matusiak and Wolf \cite{mat} have argued
that a possible reason for such significant difference
 might be that Zhang et al. \cite{ZHANG} used different samples, one for $\kappa_{xy}$ and
another one  for $\sigma_{xy}$ measurements, which makes their
results for $L_H$ inconsistent.

Actually it has been shown \cite{alelor} that there is no
inconsistency in both $L_H$ determinations. One order of magnitude
difference in two independent direct measurements of the
normal-state Hall-Lorenz number
 is consistently explained by the bipolaron theory \cite{alebook1}. The theory explains the huge difference in the
Hall-Lorenz numbers by taking into account the difference between
the in-plane resistivity of detwinned \cite{ZHANG} and twinned
\cite{mat} single crystals. It fits well the observed $L_H(T)$s and
explains a sharp Hall-number maximum \cite{mat} observed in the
normal state of underdoped cuprates.

In the presence of the electric field \textbf{E}, the temperature
gradient $\boldsymbol\nabla{T}$ and a weak magnetic field
\textbf{B}$\parallel$ \textbf{z} $\perp$ \textbf{E} and
$\boldsymbol\nabla{T}$, the electrical currents in $x,y$ directions
are  given by
\begin{eqnarray}
j_{x}\mathbf{=}a_{xx}\nabla_x(\mu-2e\phi)+a_{xy}
\nabla_y(\mu-2e\phi) \notag
\\
+b_{xx}\nabla_xT+b_{xy}\nabla_yT, \notag
\\
j_{y}\mathbf{=}a_{yy}\nabla_y(\mu-2e\phi)+a_{yx}\nabla_x(\mu-2e\phi)
\notag
\\
+b_{yy}\nabla_yT+b_{yx}\nabla_xT, \notag
\\
\end{eqnarray}
and the thermal currents are:
\begin{eqnarray}
w_{x}\mathbf{=}c_{xx}\nabla_x(\mu-2e\phi)+c_{xy}\nabla_y(\mu-2e\phi)
\notag
\\
+d_{xx}\nabla_xT+d_{xy}\nabla_yT \notag
\\
w_{y}\mathbf{=}c_{yy}\nabla_y(\mu-2e\phi)+c_{yx}\nabla_x(\mu-2e\phi)
\notag \\
 +d_{yy}\nabla_yT+d_{yx}\nabla_xT.
 \notag
 \\
\end{eqnarray}
Here $\mu$ and $\phi$ are the chemical and electric potentials.

Real phonons and (bi)polarons are well decoupled in the
strong-coupling regime of the electron-phonon interaction
\cite{alebook1} so the standard Boltzmann equation for the kinetics
of renormalised carriers is applied. If we make use of the $\tau(E)
-$approximation \cite{ANSE}  the kinetic coefficients of bipolarons
 are found as \cite{leeale}
\begin{eqnarray}
a^{b}_{xx}&=&a^{b}_{yy}=\frac{2en_b}{m_b}\langle\tau_b\rangle,\notag
\\
a^{b}_{yx}&=&-a^{b}_{xy}=\frac{2eg_bBn_b}{m_b}\langle\tau_b^2\rangle,
\notag
\\
b^{b}_{xx}&=&b^{b}_{yy}=\frac{2en_b}{Tm_b}\langle(E-\mu)\tau_b\rangle,
\notag
\\
b^{b}_{yx}&=&-b^{b}_{xy}=\frac{2eg_bBn_b}{Tm_b}\langle(E-\mu)\tau_b^2\rangle,
\notag
\end{eqnarray}
and
\begin{eqnarray}
c^{b}_{xx}&=&c^{b}_{yy}=\frac{n_b}{m_b}\langle(E+2e\phi)\tau_b\rangle,\notag
\\
c^{b}_{yx}&=&c^{b}_{xy}\frac{g_bBn_b}{m_b}\langle(E+2e\phi)\tau_b^2\rangle,
\notag
\\
d^{b}_{xx}&=&d^{b}_{yy}=\frac{n_b}{Tm_b}\langle(E+2e\phi)(E-\mu)\tau_b\rangle,\notag
\\
d^{b}_{yx}&=&-d^{b}_{xy}=\frac{g_bBn_b}{Tm_b}\langle(E+2e\phi)(E-\mu)\tau_b^2\rangle,
\notag
\end{eqnarray}
where
\begin{equation}
\langle Q(E) \rangle=\frac{\int_{0}^{\infty}dE Q(E) E
D_{b}(E)\partial f_{b}/\partial E}
{\int_{0}^{\infty}dEED_b(E)\partial f_{b}/\partial E},
\end{equation}
$D_b(E)\propto E^{d/2-1}$ is the density of  states of a
$d$-dimensional bipolaron spectrum, $E=K^2/(2m_b)$, $g_b=2e/m_b$,
and $f_b(E)$ is the equilibrium distribution function. Polaronic
coefficients are obtained by replacing super/subscripts $b$ for $p$,
double elementary charge $2e$ for $e$ and $\mu$ for $\mu/2$ in all
kinetic coefficients,
 and $m_b$ for $2m_p$ in $a_{ij}$ and $c_{ij}$.
The kinetic energy of bipolarons, $E$ should be replaced by $E+T^*$,
where $E=k^2/(2m_p)$ is the polaron kinetic energy, and $T^*$ is
half of the bipolaron binding energy (i.e. the pseudogap temperature
in the theory \cite{alebook1}).

The in-plane resistivity, $\rho$, the Hall number, $R_H$,  and the
Hall-Lorenz number, $L_H$ are expressed in terms of the kinetic
coefficients as $\rho^{-1}=2ea_{xx}$, $R_H=a_{yx}/2eB(a_{xx})^2$,
and
\begin{equation}
L_{H}=\frac{e\left[(d_{yx}a_{xx}-c_{yx}b_{xx})a_{xx}-c_{xx}(b_{xx}a_{yx}-b_{yx}a_{xx})\right]}
{2Ta_{yx}a_{xx}^2},
\end{equation}
 respectively, where $a,b,c,d=a^p+a^b,b^p+b^b,c^p+c^b,d^p+d^b$.

The $in$-plane resistivity, the temperature-dependent paramagnetic
susceptibility, and  the Hall ratio have  been  described by the
bipolaron model taking into account thermally activated single
polarons \cite{alebra,jung,in,alezavdzu}. The bipolaron model has
also offered a  simple explanation of  $c$-axis transport and the
anisotropy of cuprates \cite{alekabmot,hof2,in,zve}. The crucial
point is that single polarons dominate in  $c$-axis transport at
finite
temperatures because they are much lighter than bipolarons in  $c$%
-direction. Bipolarons can propagate across the planes only via a
simultaneous two-particle tunnelling, which is much less probable
than a single polaron tunnelling. However, along the planes polarons
and inter-site bipolarons propagate with  comparable effective
masses (section 2). Hence in the mixture of nondegenerate
quasi-two-dimensional (2D) bosons and thermally excited 3D fermions,
only fermions contribute to $c$ -axis transport, if the temperature
is not very low, which leads to the thermally activated $c$ -axis
transport and to the huge anisotropy  of cuprates \cite{alekabmot}.

We have also shown \cite{leeale} that by the necessary inclusion of
thermally activated polarons, the model predicts a breakdown of the
WF law with the small near-linear in temperature
 Hall-Lorenz number, as observed experimentally by Zhang et
al. \cite{ZHANG} (see Fig.8). Let us now show that the bipolaron
model  describes the contrasting observations of Ref. \cite{mat} as
well, if the ratio of bipolaron and polaron mobilities,
$\alpha=2\tau_b m_p/\tau_p m_b$ becomes relatively small.

 Both polaronic and bipolaronic carriers are not degenerate above
 $T_c$, so the  classical distribution functions,
 $f_b=y\exp(-E/T)$ and $f_p=y^{1/2} \exp[-(E+T^*)/T]$ are applied with
 $y=\exp(\mu/T)$. The chemical potential is evaluated
using $2n_b+n_p=x/v_0$, where $x$ is the number of itinerant holes
in the unit cell volume $v_0$  not localised by disorder. The
bipolaron density remains large compared with the polaron density in
a wide temperature range, so that  $n_bv_0\approx x/2$ and $y
\approx \pi x/(m_ba^2T)$ for quasi-2D bipolarons. Then the atomic
density of 3D polarons  is $n_pv_0=Tm_pa^2 \exp(-T^*/T)
 (xm_p/2\pi^2m_b)^{1/2}$ ($a$ is the lattice constant). The ratio $\beta=n_p/2n_b$ remains small at any
 pseudogap temperature $T^*$ and any relevant doping
 level $x > 0.05$, $\beta \approx T exp(-T^*/T)(18 m_p/\pi^2 xm_b)^{1/2}/W \ll 1$, if the temperature $T$ is
 small compared with  the polaron bandwidth $W=6/m_pa^2$. Hence, if the
 mobility ratio $\alpha$ is of the order of unity,
 both longitudinal and transverse in-plane magnetotransport is
 dominated by bipolarons, which explains a remarkably low $L_H$ in
 high-quality detwinned crystals used in Ref.\cite{ZHANG}, Fig.8.

{\begin{figure}
\begin{center}
\includegraphics[angle=-90,width=0.50\textwidth]{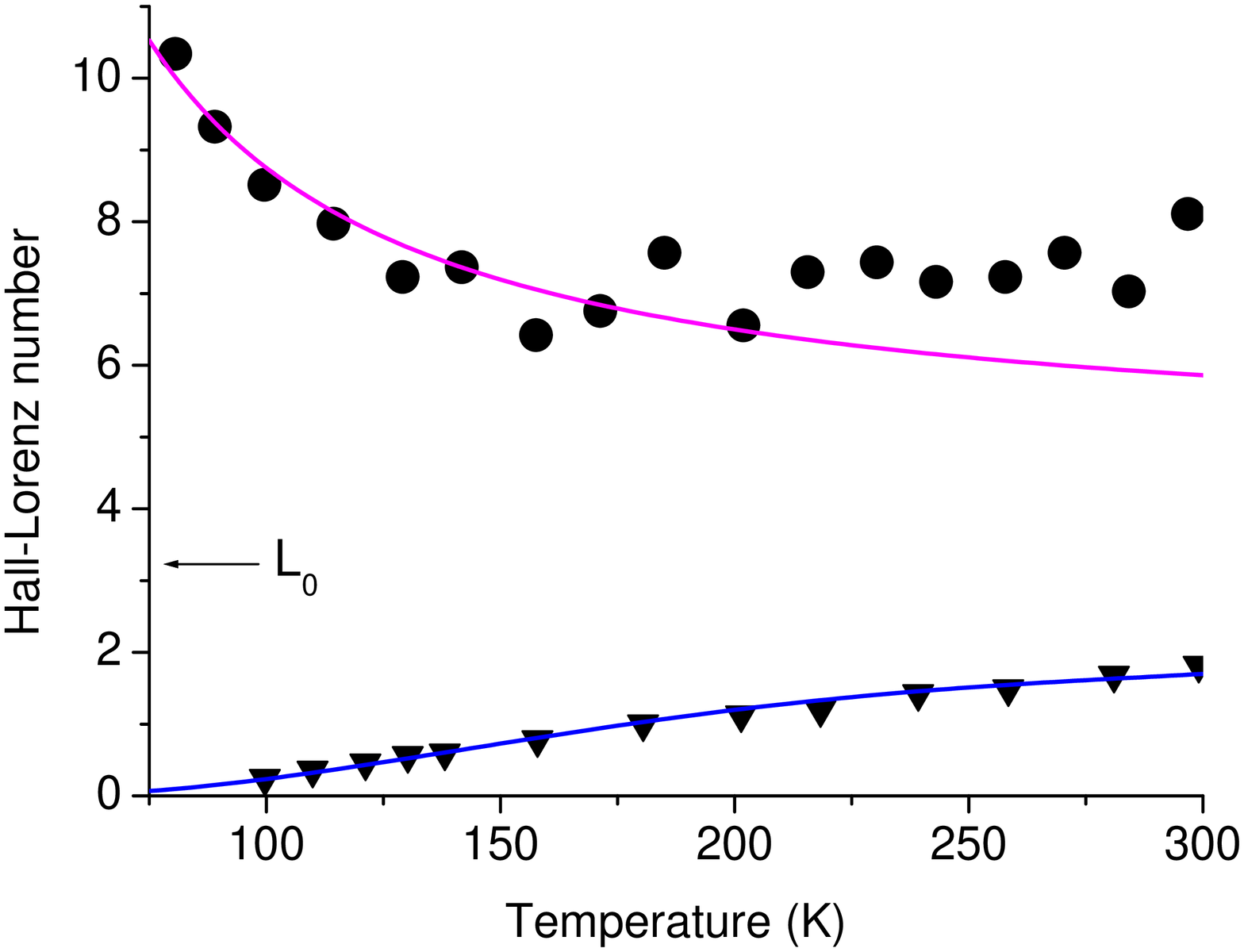}
\vskip -0.5mm \caption{The Hall-Lorenz number $L_H$ in underdoped
twinned $EuBa_{2}Cu_{3}O_{6.65}$  (circles)\cite{mat} compared with
the theory, Eq.(28) when $\alpha \ll 1$(upper line), and  the
significantly different Hall-Lorenz number in detwinned
$YBa_{2}Cu_{3}O_{6.95}$  (triangles)\cite{ZHANG} described by the
same theory \cite{leeale}  with a moderate value of
 $\alpha=0.44$ (lower line).}
\end{center}
\end{figure}

 On
 the other hand, twinned crystals used in Ref.\cite{mat} had the
 in-plane resistivity several times larger than those of Ref. \cite{ZHANG}
 presumably resulting from  twin boundaries and long
 term aging. The twin boundaries and other  defects are
 strong scatterers for slow 2D bipolarons, while lighter quasi-3D polarons
are mainly scattered by real optical phonons, which are  similar in
all crystals. Hence one can expect that $\alpha$ becomes small in
twinned crystals of Ref. \cite{mat}. If the condition $\alpha^2 \ll
\beta$ is met, then only polarons contribute to the transverse
electric and thermal magnetotransport. It explains about the same
thermal Hall conductivities ($\kappa_{xy} \approx 2.5\times 10^{-3}
B$ W/Km at T=100K) dominated by polarons in both crystals of
$YBa_{2}Cu_{3}O_{6.95}$ used in Ref. \cite{ZHANG} and in Ref.
\cite{mat},   and at the same time a substantial difference of their
electrical Hall conductivities, $\sigma_{xy}$, as bipolarons
virtually do not contribute to $\sigma_{xy}$ in the twinned samples.

To arrive at simple
 analytical results and illustrate their quantitative agreement with the experiment \cite{mat} let us  assume
 that $\alpha^2 \ll\beta$, but $\alpha \gtrsim \beta$,
 and neglect an energy dependence of the transport relaxation rates of all carriers.
  In such conditions
 bipolarons  do not contribute
to transverse heat and electric flows, but determine the in-plane
conductivity.  Kinetic responses are grossly simplified as

\begin{eqnarray}
\rho={m_bv_0\over{2e^2x  \tau_b}}
\\
 R_H={v_0\beta\over{ex\alpha^2}}={e^3 n_p  \tau_p^2\over{m_p^2}}\rho^2,
\\
L_{H}=4.75+3T^*/T+(T^*/T)^2.
\end{eqnarray}

As in the case of $\alpha^2 \gtrsim \beta$, discussed in Ref.
\cite{leeale}, the recombination of a pair of polarons into
bipolaronic bound states at the cold end of the sample  results in
the breakdown of the WF law, as described by two
temperature-dependent terms in Eq.(28). The breakdown is reminiscent
of the one  in conventional semiconductors caused by the
recombination of electron-hole pairs at the cold end \cite{ANSE}.
However, the temperature dependence and the value of $L_H(T)$ turn
out remarkably different. When $\alpha^2 \ll \beta$, The Hall-Lorenz
number is more than by an order of magnitude larger than in the
opposite regime, $\alpha^2 \gtrsim \beta$. It increases with
temperature lowering rather than decreases fitting well the
experimental observation\cite{mat} in twinned underdoped single
crystals of $EuBa_{2}Cu_{3}O_{6.65}$ with $T^*=$ 100K, Fig.8. Hence
by varying the bipolaron to polaron mobility ratio, $\alpha$, the
model accounts for qualitatively different behaviours of $L_H(T)$ in
twinned and detwinned cuprates.

\subsection{Normal state diamagnetism}

A number of experiments (see, for example,
\cite{mac,junM,hof,nau,igu,ong} and references therein), including
torque magnetometries, showed enhanced diamagnetism above $T_c$,
which has been explained as the fluctuation diamagnetism in quasi-2D
superconducting cuprates (see, for example Ref. \cite{hof}). The
data taken at relatively low magnetic fields (typically below 5
Tesla) revealed a crossing point in the magnetization $M(T,B)$ of
most anisotropic cuprates (e.g. $Bi-2212$), or in $M(T,B)/B^{1/2}$
of less anisotropic $YBCO$ \cite{junM}. The dependence of
magnetization (or $M/B^{1/2}$) on the magnetic field has been shown
to vanish at some characteristic temperature below $T_c$. However
the data taken in high magnetic fields (up to 30 Tesla) have shown
that the crossing point, anticipated for low-dimensional
superconductors and associated with superconducting fluctuations,
does not explicitly exist in magnetic fields above 5 Tesla
\cite{nau}.

Most surprisingly the torque magnetometery  \cite{mac,nau} uncovered
a diamagnetic signal somewhat above $T_c$ which increases in
magnitude with applied magnetic field. It has been  linked with the
Nernst signal and mobile vortexes   in the  normal state of cuprates
\cite{ong}. However, apart from the inconsistences mentioned above,
the vortex scenario of the normal-state diamagnetism is internally
inconsistent.  Accepting the vortex scenario and fitting  the
magnetization data in $Bi-2212$  with the conventional  logarithmic
field dependence \cite{ong}, one obtains surprisingly high upper
critical fields $H_{c2} > 120$ Tesla and a very large
Ginzburg-Landau parameter, $\kappa=\lambda/\xi >450$  even at
temperatures close to $T_c$. The in-plane low-temperature magnetic
field penetration depth is $\lambda=200$ nm in optimally doped
$Bi-2212$ (see, for example \cite{tallon}). Hence the zero
temperature coherence length $\xi$ turns out to be about  the
lattice constant, $\xi=0.45$nm, or even smaller. Such a small
coherence length rules out the "preformed Cooper pairs"  \cite{kiv},
since the pairs are virtually not overlapped at any size of the
Fermi surface in $Bi-2212$ . Moreover the magnetic field dependence
of $M(T,B)$ at and above $T_c$ is entirely inconsistent  with what
one expects from a vortex liquid.  While $-M(B)$  decreases
logarithmically at temperatures well below $T_c$, the  experimental
curves \cite{mac,nau,ong} clearly show that   $-M(B)$  increases
with the field at and  above $T_c$ , just opposite to what one could
expect in the vortex liquid.  This significant departure from the
London liquid behavior clearly indicates that the vortex liquid does
not appear above the resistive phase transition \cite{mac}.

Some time ago we  explained the anomalous diamagnetism in cuprates
as the Landau normal-state diamagnetism of preformed bosons
\cite{den}. The same model  predicted  the unusual upper critical
field \cite{aleH} observed in many superconducting cuprates
\cite{buc,mac0,boz,lawrie,gan,alezavnev}. More recently the model
has been extended   to high magnetic fields taking into account the
magnetic pair-breaking of singlet bipolarons and the anisotropy of
the energy spectrum \cite{aledia}. When the
 magnetic field is applied perpendicular to the copper-oxygen
plains the quasi-2D bipolaron energy spectrum is quantized as
$E_\alpha= \omega(n+1/2) +2t_c [1-\cos(K_zd)]$, where $\alpha$
comprises $n=0,1,2,...$ and  in-plane $K_x$ and out-of-plane $K_z$
center-of-mass quasi-momenta, $\omega=2 eB/\sqrt{m^{\ast \ast}_x
m^{\ast \ast}_y}$, $t_c$ and $d$ are the hopping integral and the
lattice period perpendicular to the planes. We assume here that the
spectrum consists of two degenerate branches, so-called $"x"$ and
$"y"$ bipolarons  as in the case of apex intersite pairs \cite{ale5}
with anisotropic in-plane bipolaron masses $m^{\ast \ast}_x\equiv m$
and $m^{\ast \ast}_y\approx 4m$. Expanding the Bose-Einstein
distribution function in powers of $\exp[(\mu-E)/T]$ with the
negative chemical potential $\mu$ one can after summation over $n$
readily obtain
 the boson density
\begin{equation}
n_b={2eB\over{\pi  d}} \sum_{r=1}^{\infty} I_0(2t_c r/T) {\exp[
(\mu-\omega/2 -2t_c)r/T]\over{1-\exp(-\omega r/T)}},
\end{equation}
and the magnetization,
\begin{eqnarray}
&&M(T,B)=-n_b \mu_b+  {2eT\over{\pi  d}} \sum_{r=1}^{\infty}
I_0\left({2t_c r\over {T}}\right)\times \\
&&{\exp[ (\mu-\omega/2 -2t_c)r/T]\over{1-\exp(- \omega r/T)}}
\left({1\over{r}}-{\omega \exp(-\omega r/T)\over{k_BT[1-\exp(-\omega
r/T)]}}\right).\nonumber
\end{eqnarray}
Here $\mu_b= e/\sqrt{m^{\ast \ast}_xm^{\ast \ast}_y}$
 and $I_0(x)$ is the modified Bessel
function. At low temperatures $T \rightarrow 0$ Schafroth's result
\cite{shaf} is recovered, $M(0,B)= -n_b \mu_b$. The magnetization of
charged bosons is field-independent at low temperatures. At high
temperatures, $T \gg T_c$ the chemical potential has a large
magnitude, and we can keep only the terms with $r=1$ in Eqs.(85,86)
to obtain $M(T,B)=-n_b \mu_b \omega/(6T)$ at $T \gg T_c\gg \omega$,
 which is the familiar  Landau  orbital diamagnetism  of nondegenerate
 carriers. Here $T_c$ is the  Bose-Einstein condensation
temperature $T_{c}= 3.31(n_{b}/2)^{2/3}/(m^{\ast \ast}_{x}m^{\ast
\ast}_{y}m^{\ast \ast}_{c})^{1/3}$, with $m_{c}=1/2|t_{c}|d^{2}$.

\begin{figure}
\begin{center}
\includegraphics[angle=-90,width=0.70\textwidth]{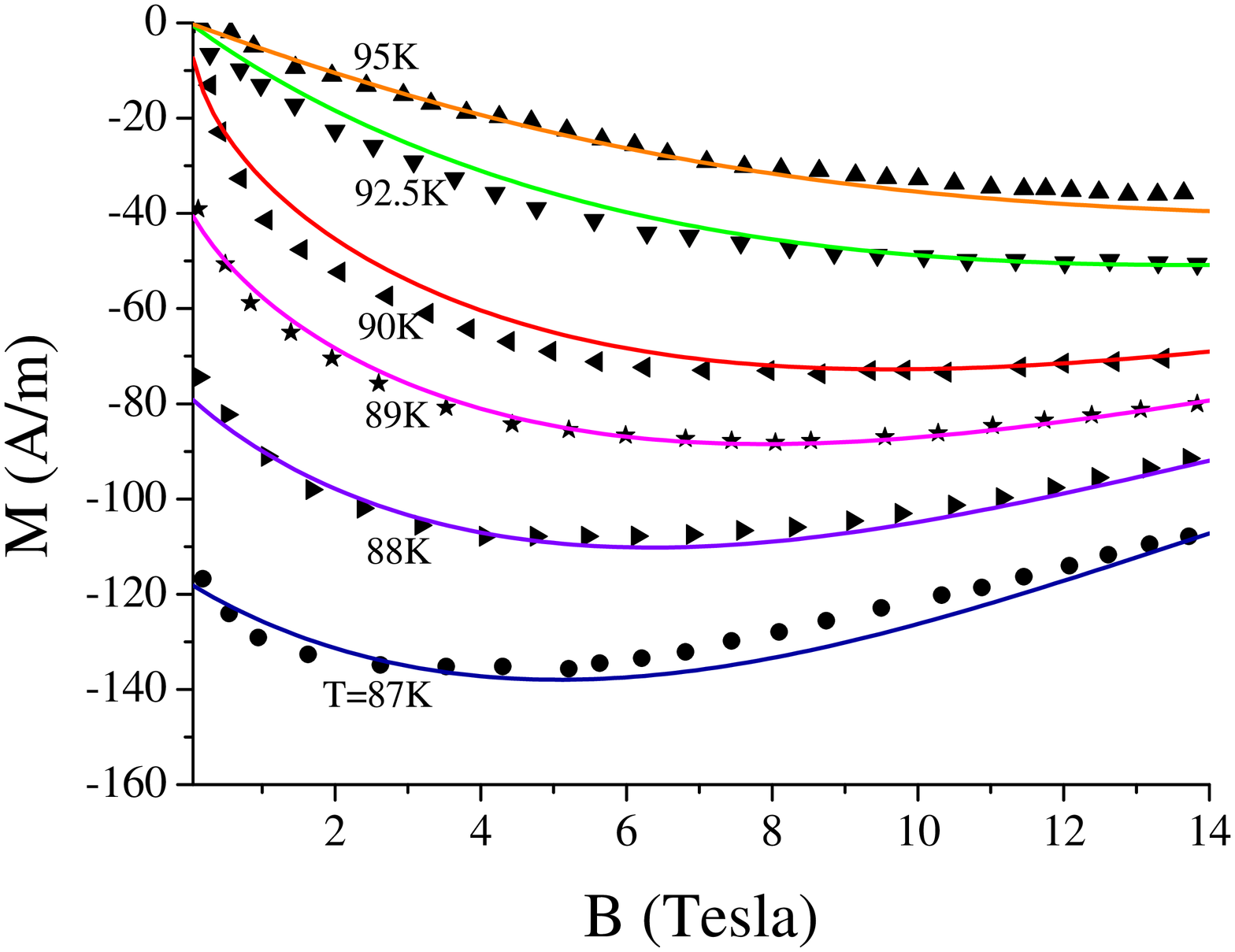}
\vskip -0.5mm \caption{Diamagnetism of optimally doped Bi-2212
(symbols)\cite{ong} compared with magnetization of CBG \cite{aledia}
near and above $T_c$ (lines).}
\end{center}
\end{figure}

Comparing with experimental data one has to take into account a
temperature and field depletion of singlets due to their thermal
excitations into spin-split triplets and single polaron states,
Fig.4. If $J <\Delta/2$,
 triplets mainly contribute to  temperature and field dependencies
of the singlet bipolaron density near $T_c$,
 $n_b(T,B)=n_c[1-\alpha \tau -(B/B^*)^2]$. Here
 $\alpha=3(2n_ct)^{-1}[J (e^{J/T_c}-1)^{-1}-T_c\ln(1-e^{-J/T_c})]$,
 $\mu_BB^*=(2T_cn_ct)^{1/2}
\sinh(J/2T_c)$, $\mu_B\approx 0.93 \times 10^{-23}$ Am$^2$ is the
 Bohr magneton, $n_c$ is the density of singlets at $T=T_c$  in zero
 field, $\tau=T/T_c-1$,
 and $2t$ is the triplet bandwidth. A triplet contribution to diamagnetism remains negligible compared with the singlet
 diamagnetism if $|1-T/T_c| \ll  J/T_c$. Also bosons localised in a random potential
contribute to the diamagnetism. However their
  contribution remains small compared with the extended carrier
 diamagnetism, if the localization energy is large compared with
 $T$. As a result,  Eq.(30) fits remarkably well the  experimental curves in the critical
region of  optimally doped Bi-2212, Fig.9,   with $n_c \mu_b=
2100$A/m, $T_c=90$K,  $\alpha=0.62$ and $B^*=56$ Tesla, which
corresponds to the singlet-triplet exchange energy $J\approx 20$K.

\subsection{Giant proximity effect}
Several groups reported that in the Josephson cuprate \emph{SNS}
junctions  supercurrent can run through normal \emph{N}-barriers
thicker than 100 nm in a strong conflict with the standard
theoretical picture, if the barrier is made from non-superconducting
cuprates. Using an advanced molecular beam epitaxy, Bozovic \emph{et
al.} \cite{bozp} proved that this giant proximity effect (GPE) is
intrinsic, rather than extrinsic caused by any inhomogeneity of the
barrier. Hence GPE defies the conventional explanation, which
predicts that the critical current should exponentially decay with
the characteristic length of about the coherence length, which is
$\xi \lesssim 1$ nm in the cuprates.

Here I show that the effect can be broadly understood as the
Bose-Einstein condensate (BEC) tunnelling into
 a cuprate
\emph{semiconductor}. As mentioned in section 1, the chemical
potential $\mu$ remains in the charge-transfer gap of doped cuprates
like La$_{2-x}$Sr${_x}$CuO$_4$ \cite{boz0}  because of  bipolaron
formation. The condensate wave function, $\psi(Z)$,  is described by
the Gross-Pitaevskii (GP) equation.   In the superconducting region,
$Z<0$, near the $SN$ boundary, Fig.10, the equation is
\begin{equation}
{1\over{2m^{**}_c}}{d^2\psi(Z)\over{dZ^2}}=[V
|\psi(Z)|^2-\mu]\psi(Z),
\end{equation}
where $V$ is a short-range repulsion of bosons, and $m^{**}_c$ is
the boson mass along $Z$.  Deep inside the superconductor
$|\psi(Z)|^2=n_s$ and $\mu=Vn_s$ , where the condensate density
$n_s$ is about $x/2$, if the temperature is well below $T_c$  of the
superconducting electrode (the in-plane lattice constant $a$ and the
unit cell volume are
 taken as unity).

\begin{figure}
\begin{center}
\includegraphics[angle=-90,width=0.70\textwidth]{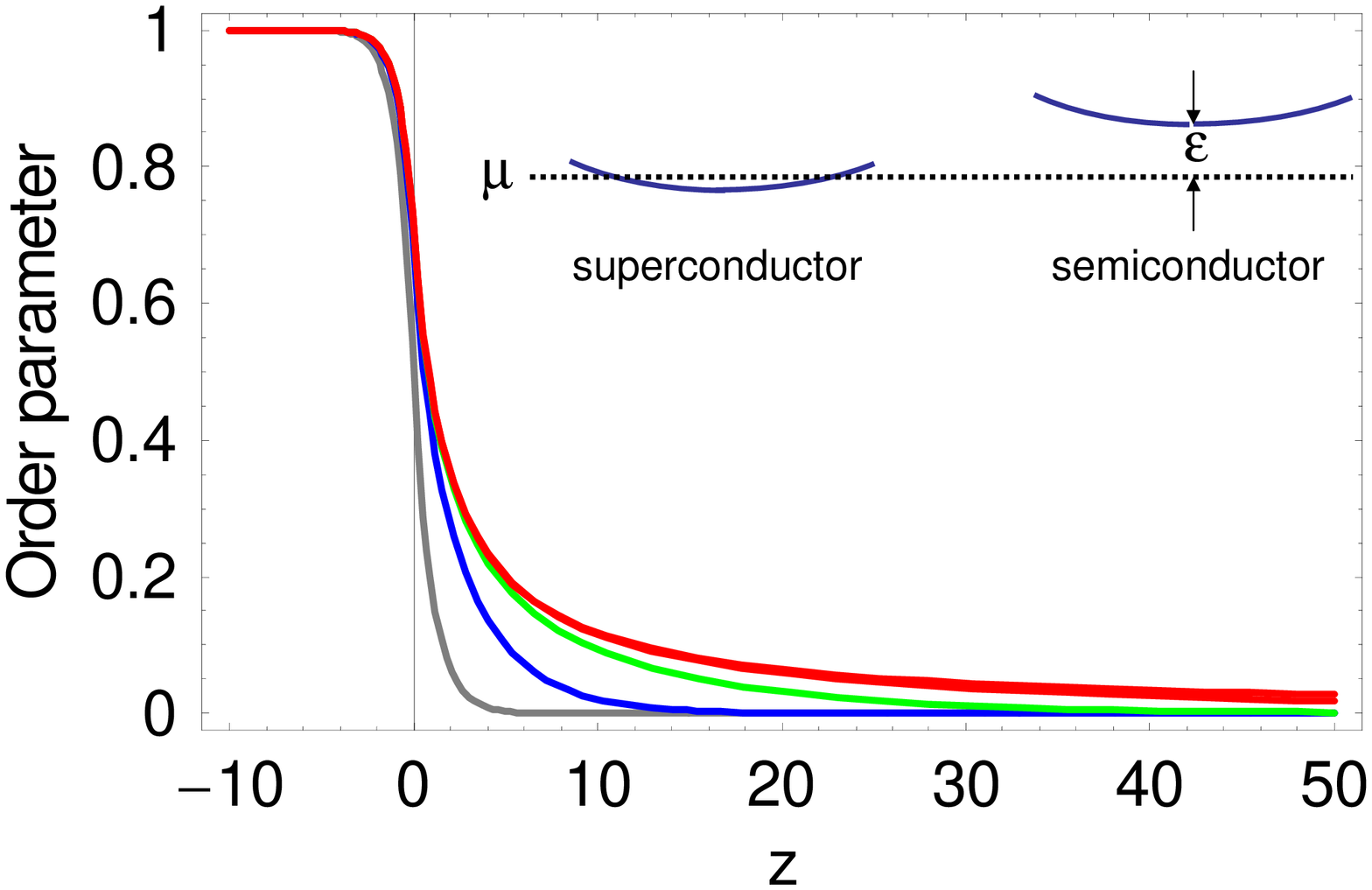}
\vskip -0.5mm \caption{BEC order parameter at the $SN$ boundary for
$\tilde{\mu}=1.0,0.1,0.01$ and $ \leqslant 0.001$ (upper curve).}
\end{center}
\end{figure}

The normal barrier  at $Z
>0$ is an underdoped cuprate semiconductor above its
transition temperature, where the chemical potential $\mu$ lies
below the bosonic band by some energy $\epsilon$, Fig.10.  One
obtains for quasi-two dimensional bosons \cite{alebook1}
\begin{equation}
\epsilon=-T\ln(1-e^{-T_0/T}),
\end{equation}
where $T_0=\pi x'/m^{**}$, $m^{**}$ is the in-plane boson mass, and
$x' < x$ is the doping level of the barrier. Then the GP equation in
the barrier can be written as
\begin{equation}
{1\over{2m^{**}_c}}{d^2\psi(Z)\over{dZ^2}} =[V
|\psi(Z)|^2+\epsilon]\psi(Z).
\end{equation}
Introducing the bulk coherence length, $\xi= 1/(2m^{**}_c
n_sV)^{1/2}$ and dimensionless $f(z)=\psi(Z)/n_s^{1/2}$,
$\tilde{\mu}=\epsilon/n_sV$, and $z=Z/\xi$ one obtains
 for a real
$f(z)$
\begin{equation}
{d^2f\over{dz^2}} =f^3-f,
\end{equation}
if $z<0$, and
\begin{equation}
{d^2f\over{dz^2}}=f^3+\tilde{\mu}f,
\end{equation}
if
 $z>0$. These equations can be readily solved using  first
integrals of motion respecting the boundary conditions,
$f(-\infty)=1$, and $f(\infty)=0$,
\begin{equation}
{df\over{dz}}= -(1/2+f^4/2-f^2)^{1/2},
\end{equation}
and
\begin{equation}
{df\over{dz}}= -(\tilde{\mu}f^2+f^4/2)^{1/2},
\end{equation}
for $z<0$ and $z>0$, respectively. The solution in the
superconducting electrode is given by
\begin{equation}
f(z)=\tanh \left[-2^{-1/2}z+0.5
\ln{{2^{1/2}(1+\tilde{\mu})^{1/2}+1}\over{2^{1/2}(1+\tilde{\mu})^{1/2}-1}}\right].
\end{equation}
It decays  in the close vicinity of the barrier from 1 to
$f(0)=[2(1+\tilde{\mu})]^{-1/2}$ in the interval about the coherence
length $\xi$. On the other side  of the boundary, $z>0$, it is given
by
\begin{equation}
f(z)={(2\tilde{\mu})^{1/2}\over{\sinh\{z\tilde{\mu}^{1/2}+\ln[2(\tilde{\mu}(1+\tilde{\mu}))^{1/2}+(1+4\tilde{\mu}(1+\tilde{\mu}))^{1/2}]\}}}
.
\end{equation}
Its profile is shown in Fig.10. Remarkably, the order parameter
penetrates into the normal layer up to the length $Z^* \thickapprox
(\tilde{\mu})^{-1/2}\xi$, which could be larger than $\xi$ by many
orders of magnitude,  if $\tilde{\mu}$ is  small. It is indeed the
case, if the barrier layer is sufficiently doped. For example,
taking $x'=0.1$,   c-axis $m^{**}_c=2000 m_e$, in-plane $m^{**}=10
m_e$ \cite{alebook1}, $a=0.4$ nm, and $\xi=0.6$ nm, yields
 $T_0\approx 140$ K and $(\tilde{\mu})^{-1/2}\approx 5000$ at $T=10$K. Hence the
 order parameter could penetrate into the normal cuprate semiconductor
 up to more than a thousand coherence lengths as observed \cite{bozp}. If the thickness of the barrier $L$ is small compared with $Z^*$,
and $(\tilde{\mu})^{1/2}\ll 1$, the order parameter decays following
 the power law, rather than exponentially,
\begin{equation}
f(z)={\sqrt{2}\over{z+2}}.
\end{equation}
Hence, for $L \lesssim Z^*$, the critical current should also decay
following the power law \cite{alepro}. On the other hand, for an
\emph{undoped}
 barrier $\tilde{\mu}$ becomes
 larger than unity, $\tilde{\mu}\varpropto \ln(m^{**}T/\pi x')\rightarrow \infty$ for any finite temperature $T$  when $x' \rightarrow
 0$, and the current should exponentially decay with the characteristic length  smaller that $\xi$, as is experimentally observed as well \cite{boz0}.

\section{Discussion}

A possibility of real-space pairing, as opposed to the Cooper
pairing, has been the subject of many discussions,  particularly
heated over the last 20 years after the discovery of high
temperature superconductivity in cuprates. The first proposal for
high temperature superconductivity, made by Ogg Jr in 1946
\cite{ogg}, already involved  real-space pairing of individual
electrons into bosonic molecules  with zero total spin. This idea
was further developed as a natural explanation of conventional
superconductivity by Schafroth  and Butler and Blatt \cite{shaf}.
However, with one or two exceptions, the Ogg-Schafroth picture was
condemned and practically forgotten because it neither accounted
quantitatively for the critical behavior of conventional
superconductors, nor did it explain the microscopic nature of
attractive forces which could overcome the Coulomb repulsion between
two electrons  constituting a
 pair. The failure of the `bosonic' picture of individual
electron pairs became fully transparent when Bardeen, Cooper and
Schrieffer \cite{bcs} proposed that two electrons in a
superconductor were indeed  correlated, but at a very large distance
of about $10^{3}$ times of the average inter-electron spacing.

Highly successful for low-Tc metals and alloys  the BCS theory has
led many researchers to believe that novel high-temperature
superconductors should also be  "BCS-like". However, the
Ogg-Schafroth and the BCS descriptions are actually two opposite
extremes of the same electron-phonon (e-ph) interaction. Indeed by
extending the BCS theory towards the strong interaction between
electrons and ion vibrations, a charged Bose gas (CBG) of tightly
bound  small bipolarons was predicted by us \cite{aleran} with a
further prediction  that high $T_c$ should exist in the crossover
region of the e-ph interaction strength from the BCS-like to
bipolaronic superconductivity \cite{ale0}.

However, for very strong electron-phonon coupling, polarons become
self-trapped on a single lattice site. The energy of the resulting
small polaron is given as $E_p=-\lambda zT(a)$, where $\lambda$ is
the electron-phonon coupling constant, $T(a)$ is the hopping
parameter and $z$ is the coordination number. Expanding about the
atomic limit in small $T(a)$ (which is small compared to $E_p$ in
the small polaron regime, $\lambda>1$)
 the polaron mass is
computed as $m^{*}=m_0\exp(\gamma z\lambda/\hbar\omega)$ , where
$\omega$ is the frequency of Einstein phonons, $m_0$ is the rigid
band mass on a cubic lattice, and $\gamma$ is a numerical constant.
For the Holstein model, which is purely site local, $\gamma=1$.
Bipolarons are on-site singlets in the Holstein model and their mass
$m_{H}^{**}$ appears only in the second order of $T(a)$
\cite{aleran} scaling as $m_{H}^{**}\propto (m^{*})^2$ in the limit
$\hbar\omega\gg\Delta$ , and as $m_{H}^{**}\propto(m^*)^{4}$ in a
more realistic regime $\hbar\omega\ll\Delta$ \cite{alekab0}. Here
$\Delta=2E_p-U$ is the bipolaron binding energy, and $U$ is the
on-site (Hubbard) repulsion. Since the Hubbard $U$ is about 1 eV or
larger in strongly correlated materials, the electron-phonon
coupling must be large to stabilize on-site bipolarons and the
Holstein bipolaron mass appears very large, $m_{H}^{**}/m_0>1000$,
for realistic values of phonon frequency.

 This estimate led some authors to the conclusion that the formation of
itinerant small polarons and bipolarons in real materials is
unlikely \cite{mel}, and high-temperature bipolaronic
superconductivity is impossible \cite{and2}. However, one should
note that the Holstein model is an extreme polaron model, and
typically yields the highest possible value of the (bi)polaron mass
in the strong coupling limit.  Many advanced materials with low
density of free carriers and poor mobility (at least in one
direction) are characterized by poor screening of high-frequency
optical phonons and are more appropriately described by the
long-range Fr\"ohlich electron-phonon interaction \cite{ale5}. For
this interaction the parameter $\gamma$ is less than 1
($\gamma\approx 0.3$ on the square lattice and $\gamma\approx 0.2$
on the triangular lattice), reflecting the fact that in a hopping
event the lattice deformation is partially pre-existent. Hence the
unscreened Fr\"ohlich electron-phonon interaction provides
relatively light small polarons, which are several orders of
magnitude lighter than small Holstein polarons, which is now
confirmed by several numerical studies as discussed in section 2.

This unscreened Fr\"ohlich  interaction combined with on-site
repulsive correlations can  bind holes into superlight intersite
 bipolarons (section 2). Experimental evidence for  exceptionally strong e-ph interactions
 is now so overwhelming that the bipolaronic charged bose gas
\cite{alebook1} could be a feasible alternative to  the BCS-like
scenarios of cuprates. While
some authors \cite{kiv} have dismissed any real-space pairing, advocating  a \textit{%
collective} pairing into incoherent Cooper pairs in the momentum
space at some high temperature $T^{\ast}> T_c$, I argue that the
most likely scenario is a true 3D Bose-Einstein condensation at
$T_c$ of real-space bipolarons. Our bipolaron theory predicted such
key features of cuprate superconductors as anomalous upper critical
fields, spin and charge pseudogaps, and anomalous isotope effects
 later discovered experimentally. The theory explained normal
state kinetics, high $T_c$ values, and specific heat anomalies of
cuprates (see \cite{alebook1} and references therein).

Here I have reviewed  normal-state diamagnetism, the Nernst, thermal
Hall and giant proximity effects as strong evidence for real-space
pairing and 3D BEC in cuprates.

  I thank  A.F. Andreev, I. Bozovic, L.P.
Gor'kov, J.P. Hague, D. Mihailovic, V.V. Kabanov, P.E.
  Kornilovitch, and
J.H. Samson  for illuminating discussions.  The work was supported
by EPSRC (UK) (grant no. EP/C518365/1).

\label{lastpage-01}

\end{document}